\documentclass[%
 prb,reprint,
 amsmath,amssymb,
floatfix,
]{revtex4-1}

\DeclareUnicodeCharacter{2060}{\nolinebreak}
\usepackage{graphicx}
\usepackage{dcolumn}
\usepackage{bm}
\usepackage{multirow}%
\usepackage{algorithm}%
\usepackage{algorithmicx}%
\usepackage{algpseudocode}%
\usepackage{listings}%
\usepackage{placeins}%
\usepackage{xr}%
\usepackage{lmodern}%
\usepackage{booktabs}%

\newcommand{\beginsupplement}{%
  \renewcommand{\thetable}{S\arabic{table}}%
  \renewcommand{\thefigure}{S\arabic{figure}}%
  \renewcommand{\thesection}{S\arabic{section}}%
  \renewcommand{\thesubsection}{S\arabic{section}.\arabic{subsection}}%
}
\begin{document}

\title[Superconductivity in Substitutional Ga-Hyperdoped Ge Epitaxial Thin Films]{Superconductivity in Substitutional Ga-Hyperdoped Ge Epitaxial Thin Films}

\author{Julian~A.~Steele$^{1,2,*,\dagger}$}
\author{Patrick~J.~Strohbeen$^{3,*}$}
\author{Carla~Verdi$^{1}$}
\author{Ardeshir~Baktash$^{2,4}$}
\author{Alisa~Danilenko$^{3}$}
\author{Yi-Hsun~Chen$^{1}$}
\author{Jechiel~van~Dijk$^{3}$}
\author{Frederik~H.~Knudsen$^{3}$}
\author{Axel~Leblanc$^{3}$}
\author{David~Perconte$^{3}$}
\author{Lianzhou~Wang$^{2,4}$}
\author{Eugene~Demler$^{5}$}
\author{Salva~Salmani-Rezaie$^{6}$}
\author{Peter~Jacobson$^{6,\dagger}$}
\author{Javad~Shabani$^{3,\dagger}$}

\affiliation{
$^{1}$School of Mathematics and Physics, The University of Queensland, Brisbane, 4072, QLD, Australia\\
$^{2}$Australian Institute for Bioengineering and Nanotechnology, The University of Queensland, St. Lucia, 4072, QLD, Australia\\
$^{3}$Center for Quantum Information Physics, New York University, New York, 10003, NY, USA\\
$^{4}$School of Chemical Engineering, The University of Queensland, St. Lucia, 4072,QLD, Australia\\
$^{5}$Institute for Theoretical Physics, ETH Z\"{u}rich, Z\"{u}rich, 8093, Switzerland\\
$^{6}$Department of Materials Science and Engineering, The Ohio State University, Columbus, 43210, OH, USA\\
$^{*}$These authors contributed equally to this work.\\
$^{\dagger}$Corresponding Authors: julian.steele@uq.edu.au, p.jacobson@uq.edu.au, jshabani@nyu.edu
}



\begin{abstract}
Doping-induced superconductivity in group IV elements may enable quantum functionalities in material systems accessible with well-established semiconductor technologies. Non-equilibrium hyperdoping of group III atoms into C, Si, or Ge can yield superconductivity; however, its origin is obscured by structural disorder and dopant clustering. Here, we report the epitaxial growth of hyperdoped Ga:Ge films and trilayer heterostructures by molecular beam epitaxy with extreme hole concentrations ($n_\textup{h} = 4.15 \times 10^{21}$~cm$^{-3}$, ~17.9\% Ga substitution) that yield superconductivity with a critical temperature of $T_{\textup{c}} = 3.5$~K. Synchrotron-based X-ray absorption and scattering methods reveal that Ga dopants are substitutionally incorporated within the Ge lattice, introducing a tetragonal distortion to the crystal unit cell.  Our findings, corroborated by first-principles calculations, suggest that the structural order of Ga dopants creates a narrow band for the emergence of superconductivity in Ge, establishing hyperdoped Ga:Ge as a low-disorder, epitaxial superconductor-semiconductor platform.
\end{abstract}

\maketitle

\noindent

Realization of superconductivity in group IV elements has been an enticing field of study over the past few decades for their promising application in superconducting electronics~\cite{ekimov2004bcsuper, bustarret2006sib, blase2009grpivsc, chiodi2017sijj} such as superconducting quantum bits or cryogenic CMOS control circuitry. However, difficulties such as dopant segregation and precipitation, nanocrystal formation, incoherent interfaces, and poor layer thickness control impede their integration into a new generation of quantum devices. As a prototypical semiconductor and foundry-ready material, germanium is a key component of devices that combine macroscopic superconducting coherence with the microscopic degrees of freedom in semiconductors~\cite{scappucci2021geinfo, sammak2019geqwplay, tosato2023gehardgap, schiela2024jjprog}. Historically, superconductivity in germanium has been accomplished through extreme non-equilibrium growth techniques such as ion implantation and subsequent flash annealing~\cite{herrmannsdorfer2009gage, prucnal2019, sardashti2021}. However, recent advancements in thin film epitaxy of heavily doped germanium thin films have shown great promise in achieving well-ordered, single-crystalline thin films of superconducting germanium~\cite{strohbeen2023superge}. Homoepitaxy of superconducting group IV materials allows for expanding upon the existing repertoire of superconductor-semiconductor (S-Sm) planar devices~\cite{shabani2016inas, krogstrup2015ssmepi} to enable full S-Sm-S heterostructures that have remained challenging to realize, thus limiting their applicability in quantum devices~\cite{zhao2020mergemon, shim2014bottomup}. This difficulty has renewed interest in doping-based superconductivity in group IV materials, a concept explored theoretically by Cohen in the 1960s~\cite{RevModPhys.36.240}, as recent efforts to induce superconductivity through proximity effects~\cite{tosato2023gehardgap, schiela2024jjprog} and alloying~\cite{strohbeen2024tage} have met with mixed results. This route sidesteps disordered interfaces between dissimilar materials by inducing superconductivity within an epitaxial semiconductor thin film~\cite{shim2014bottomup}, offering a unique and promising solution to the problems of material disorder and device scale-up.\\

Hyperdoping in germanium induces a superconducting state, yet routes to high quality, device-ready films have remained out of reach. Multiple studies have confirmed the existence of superconductivity~\cite{herrmannsdorfer2009gage, prucnal2019, sardashti2021, strohbeen2023superge}, however, these materials suffer from structural inhomogeneities, dopant clustering, and poor thickness control, which limit their applications. Specifically, doping at concentrations above the standard miscibility limits raises concerns over the structural stability, site disorder, and nature of the superconducting phase~\cite{sardashti2021}. Recent efforts in Ge have pushed dopant concentrations above miscibility limits~\cite{strohbeen2023superge, trumbore1960, sardashti2021, prucnal2019, herrmannsdorfer2009gage}, but maintaining coherent epitaxy and low defect densities remains challenging as most doping strategies rely on high-energy techniques such as ion implantation and flash annealing. These approaches demonstrate superconductivity but the formation of lattice defects and broad dopant profiles complicate device integration and dopant uniformity. Furthermore, questions persist about whether these hyperdoped films are intrinsically superconducting, or if Ga clusters and unintended phases are responsible for superconductivity. These uncertainties highlight the need for growth strategies more readily amenable to achieve the quality necessary for quantum devices as well as better understanding of dopant incorporation in these extreme doping regimes.\\

Here, we use molecular beam epitaxy (MBE) to grow Ga-hyperdoped Ge films, overcoming challenges associated with high dopant concentrations while maintaining coherent epitaxial growth. Crucially, better control over parasitic heating during film growth enables the formation of a smooth floating layer of Ga adatoms that promotes low-temperature growth of Ge, preserving smooth surfaces and single-crystallinity~\cite{kesan_dopant_1991}. By re-configuring the growth chamber and optimizing growth temperatures, we obtain drastic improvements in surface morphology, atomic ordering, and domain structure. We demonstrate the improved material quality in the growth of trilayer heterostructures as a proof-of-principle design for vertical Josephson junction device architectures that alleviate concerns of material disorder related to typical amorphous oxide tunnel junctions. For films with an expected Ga doping concentration of 17.9\%, we observe a superconducting transition at 3.5~K and a carrier concentration of $4.15\times10^{21}$ holes~cm$^{-3}$. Short- and long-range structural characterization methods show that Ga substitution is energetically favored within the Ge lattice and drives a tetragonal distortion reminiscent of the $\beta$-Sn phase. Density functional theory (DFT) calculations reveal a pronounced shift of the Fermi level into the valence band and a flattening of electronic bands near the R-point of the cubic Brillouin zone. Our findings establish that superconductivity originates intrinsically in Ge from substitutional Ga doping and flat- or narrow-band physics, rather than defect-mediated superconducting weak links.\\

\begin{figure*}[htbp!]
    \centering
    \includegraphics[width=0.75\linewidth]{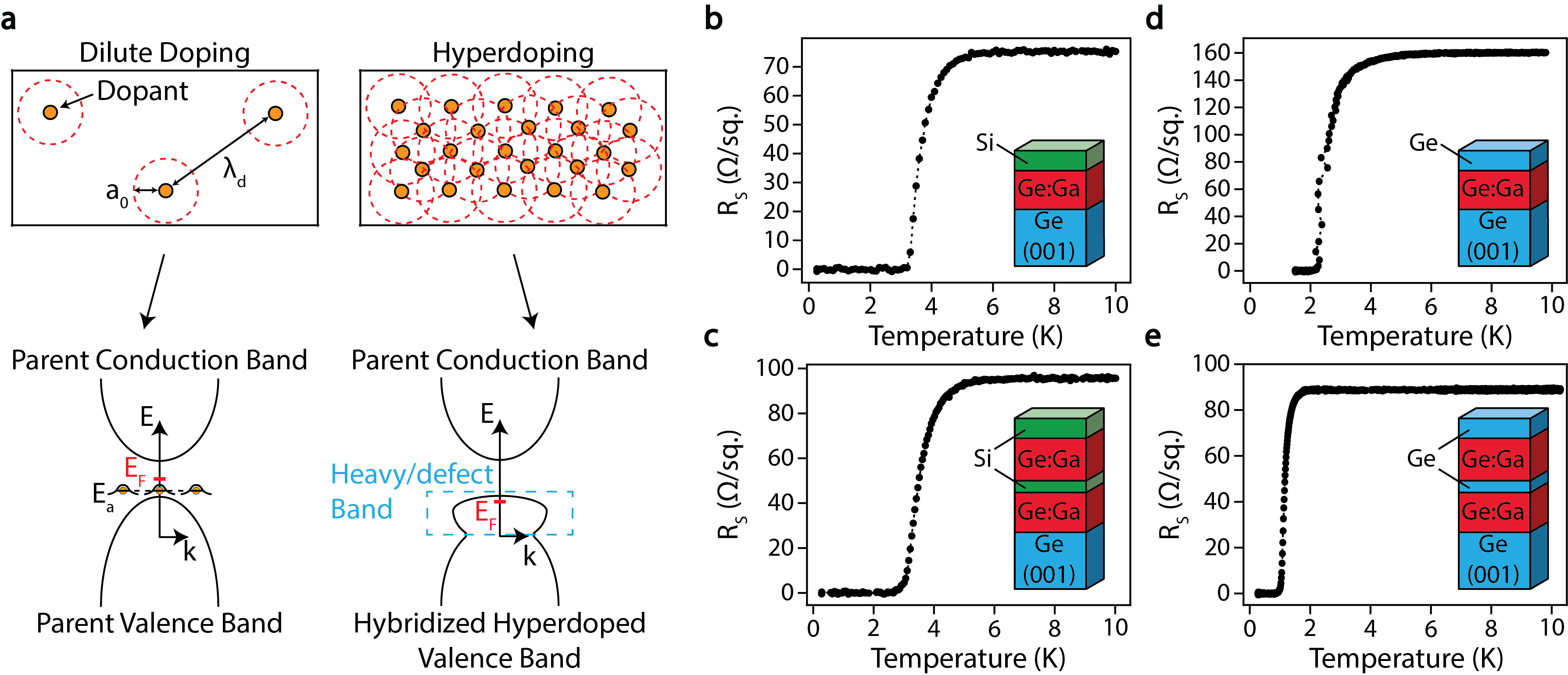}
    \caption{\textbf{Superconductivity in germanium by p-type hyperdoping.} \textbf{a} Schematic picture of dopants within a matrix and zone-center $E(k)$ dispersion for a prototypical band insulator in a dilute doping and hyperdoped state. In the dilute doping scenario, dopants with an effective Bohr radius, $a_{0}$, at an average distance between dopants, $\lambda_{d}$, induce mid-gap electronic states at an energy $E_{a}$ above the valence band edge. Increasing dopant concentrations to few-percent levels or higher induces significant hybridization between dopants, giving rise to a low-energy band associated with dopant-dopant hybridization. This defect band is traditionally expected to be non-dispersive due to the stochastic behavior of dopants. However, dopant-dopant hybridization in principle can induce ordering of the dopants on the lattice, giving rise to a narrow heavy band as well. \textbf{b-e} Sheet resistance versus temperature traces for single Ga:Ge layers and Ga:Ge sandwich structures, akin to a Josephson junction along the growth direction. In these structures, the single Ga:Ge layers are held to a constant thickness of roughly 12~nm and the trilayer structures have Ga:Ge layer thicknesses of 10~nm. The insets of each panel present schematic drawings of the sample layer structure. \textbf{b, c} are representative traces where we use Si as the cap and barrier layer material, and \textbf{d, e} are representative traces where Ge replaces the Si-containing layers. The jitter in temperature observed in \textbf{d} during the superconducting transition is an artifact caused by temperature instability during the measurement.}
    \label{fig1}
\end{figure*}

\section*{Thin film growth and material stability}

Figure~1a presents a schematic diagram of the dopant structure under dilute doping conditions, as well as a generalized picture for hyperdoping in a semiconductor, along with their respective schematic $E(k)$ dispersions. For dilute doping, the average distance between Ga dopants, $\lambda_{d}$, is larger than the effective Bohr radius of the Ga dopant, $a_{0}$. This results in discrete mid-gap states associated with these dopants at an energy $E_\textup{a}$ above the valence band edge, as is expected for a hole-doped semiconductor. However, the scenario in which dopant concentrations greatly exceed common doping levels, known as ``hyperdoping'', leads to significant dopant-dopant and dopant-semiconductor hybridization of the low-energy states. This gives rise to what is typically considered to be a non-dispersive band of defect states. Alternatively, hybridization with the host matrix can indeed give rise to a new heavy electronic band that disperses in $k$. Transport data for MBE-grown samples for this study are shown in Figure 1b and c, for structures containing a single layer of hyperdoped Ge and an epitaxial Josephson junction-like structure, respectively. Compared to previous reports on MBE-grown samples~\cite{strohbeen2023superge}, these films exhibit a slight reduction in superconducting $T_\textup{c}$, but larger carrier concentrations that are more comparable to ion implantation strategies~\cite{herrmannsdorfer2009gage, prucnal2019, sardashti2021} (see Figure~S1a). Figure~S1a shows a rough trend in $T_\textup{c}$ as a function of hole carriers where superconductivity appears near $1\times10^{21}$~cm$^{-3}$, reaching a maximum near $4-5 \times 10^{21}$~cm$^{-3}$ in the samples grown for this study. This is consistent with a previous report in boron-doped silicon~\cite{nath2024lasersib} that demonstrates an increasing hole density increases the superconducting $T_\textup{c}$. Furthermore, Figure~1b and c present the sheet resistance as a function of temperature for a 12~nm thick film of hyperdoped Ge and a trilayer heterostructure, respectively. Representative reflection high-energy electron diffraction (RHEED) images for a layered heterostructure of hyperdoped Ge films are presented in Figure~S2. During Ga doping, we observe the formation of the expected floating surface layer of Ga adatoms~\cite{kesan_dopant_1991} with an underlying single crystalline pattern that reflects the surface structure of the hyperdoped film. Further details regarding RHEED monitoring and film growth are presented in the Supplementary information and Methods, respectively. The smoothness of the resultant film surface is confirmed with \textit{ex-situ} atomic force microscopy (AFM) imaging, seen in Figure~S1b.\\

To investigate the role of the silicon interlayer more explicitly, films grown without silicon were also studied. The results of which are presented in Figures~1d, e, and~S3. We note that the silicon-containing samples stabilize slightly higher carrier concentrations, which we expect to exist due to the band offset between the hyperdoped Ge and strained Si layers~\cite{monch1987, vandewalle1988}. Silicon incorporation also enhances the resultant superconducting $T_\textup{c}$, which we attribute to a behavior akin to the BCS isotope effect~\cite{blase2009grpivsc, nath2024lasersib}. Temperature stability of the superconducting phase is shown in Figure~S3, where we obtain a maximal superconducting transition temperature at growth temperatures ranging from 100-150$^{\circ}$C. Above 150$^{\circ}$C the superconducting phase weakens as observed by a reduction in $T_\textup{c}$, coinciding with a reduction in hole carriers seen in Figure~S4. We further investigate the robustness of the superconducting phase by fabricating a Hall bar device out of the sample shown in Figure~1c. We present the data for this test device in Figure~S5 and the fabrication procedure is found in the Methods. We note no significant change in superconducting $T_\textup{c}$, out-of-plane $H_\textup{c}$, nor carrier concentration. An extended discussion of this data is provided in the Supplementary Information. Thus, we conclude that this superconducting phase is demonstrably stable during processing and suitable for device applications.\\

\begin{figure*}[htbp!]
    \centering
    \includegraphics[width=0.75\linewidth]{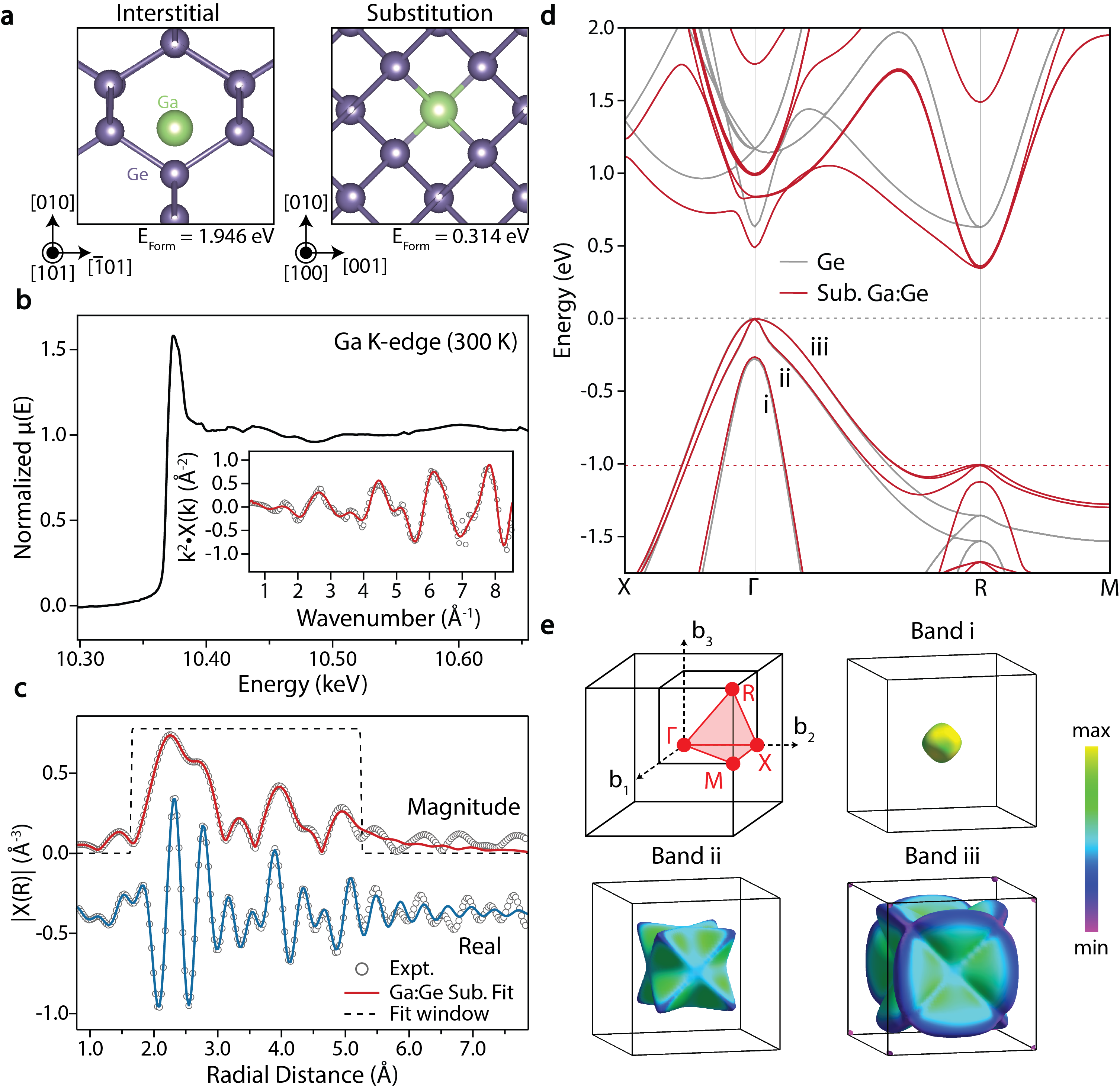}
    \caption{\textbf{Ga dopant arrangement in Ge and modulated band structure.} \textbf{a} Optimized DFT structures of Ga-doped Ge at the interstitial and substitutional sites, along with the defect formation energies ($E_\textup{Form}$).  \textbf{b} Experimental Ga-K-edge spectrum recorded from the Ga:Ge with its corresponding $k^{2}$-weighted $\chi(k)$ EXAFS fit (red line) contained in the inset. \textbf{c} The Fourier-transform ($k^{2}$-weighted) of the EXAFS data (distances have been phase-corrected). An agreeable fit has been made to the experiment using the substitutional doped crystallographic information file (CIF) generated from the optimized DFT calculated structure shown in \textbf{a}.\textbf{d} Calculated band structure for pristine Ge and the substitutionally doped GaGe$_{7}$ structure. The horizontal dashed red line indicates the position of the Fermi level after doping, shifting 1.01~eV into the valence band. \textbf{e} The contribution of each band identified as i, ii, and iii to the 3D Fermi surface with the color map corresponding to the Fermi velocity. The predicted narrow-band condition is observed as small pockets of high-mass carriers in the surface of Band iii at the corners of the Brillouin Zone (R point).}
    \label{fig2}
\end{figure*}

\section*{Dopant arrangement and band structure}

Approximating from our Hall measurements and MBE growth conditions, doping concentrations of 12.5\% and 11.1\% are simulated for the case of substitutional and interstitial doping, respectively, as detailed in the Methods section. Figure~2a shows the DFT-optimized structures. We find positive defect formation energies for cells containing Ga located at both doping sites (Figure~2a and Table~S1), implying some degree of non-equilibrium growth necessary (i.e., MBE growth) in both cases. However, the formation energy of the substitutional Ga-doped structure (0.314~eV) is markedly lower than the interstitial positions (1.946~eV), indicating that substitutional dopants are expected to be more energetically favorable in the material. This preference is inherently connected to the requirement of the Ge host to undergo an expansion to accommodate the larger Ga atom at the interstitial site, as reported in Table~S1. From our DFT calculations, substituting Ge with Ga at these doping levels is expected to impart a relatively small out-of-plane expansion of the lattice parameter of $\sim$0.12\%, resulting in a slight tetragonal distortion to the structure (Table~S1) to accommodate the heterovalent group-III dopant in a lattice-matched film.\\

To verify our theoretical calculations of doping site preference in hyperdoped Ge, we perform synchrotron X-ray absorption spectroscopy measurements which have previously demonstrated strong capabilities for determining the lattice position of dopant species~\cite{zhang2024zndopant}. From our calculations (Figure~S6), X-ray absorption at the Ga K-edge is inherently sensitive to the Ga-related local fine structure and can be used to unambiguously resolve its location within the lattice. Experimentally, we employ extended X-ray absorption fine structure (EXAFS) analysis, as measured from the Ga:Ge thin film in a fluorescent mode. X-ray absorption data is shown in Figure~2b measured from various points across the sample, normalized and then averaged together with the $k^{2}$-weighted EXAFS signal shown in the inset. The corresponding analysis of the Fourier transform of this data is presented in Figure~2c. The EXAFS data are best described by a substitutional doping model the strongest Ge-Ga nearest-neighbor scattering pair at 2.447~\r{A}, in excellent agreement with the calculated bond length of 2.439~\r{A} (Figure~2a). To rule out other viable dopant coordination environments, such as $\alpha$-Ga segregation and the interstitial lattice position, we simulate the expected profile for these alternative fine-structure environments, shown in Figure~S7. These simulations, in combination with no significant fine structure observed between 2.5 and 3.8~\r{A} in the experimental Fourier transform, we confirm that the Ga dopants preferentially occupy substitutional lattice positions, in agreement with our ab initio calculations. Furthermore, the relatively large Debye–Waller factors ($\sigma^{2}$(~\r{A}$^{2}$)) derived from the EXAFS fits (Table~S2) highlight the role of both lattice strain and Si diffusion in the Ga:Ge layers.\\

Using the optimized structures in Figure~2a with the known dopant arrangement, we now study how the electronic structure is perturbed through large incorporations of heterovalent Ga dopants. While a slight tetragonal distortion is anticipated in strained material due to substrate clamping, we find that the calculated band structure is insensitive to this distortion and we use the high-symmetry cubic Brillouin zone and paths for simplicity. The $k$-resolved band structure is presented in Figure~2d and the $k$-integrated DOS is in Figure~S8a. An evaluation of the DOS projected onto different atomic orbitals in Figure~S8b shows that new, localized states emerge through the hybridization of both the Ga and Ge p-orbitals. As expected, Ga substitution induces p-type doping in the system, with the Fermi level shifting 1.01~eV into the valence band (see Figure~2d). This corresponds to a hole carrier concentration of $5.5\times10^{21}$~cm$^{-3}$ that closely approximates the hole concentration from our Hall measurements. Furthermore, due to the lifting and flattening of the bands near the R point of the cubic Brillouin zone, we predict this system to be a candidate narrow-band superconductor. By plotting the Fermi surface and separating the contribution of each of the three bands crossing the Fermi level  (band i, ii, and iii), we observe in the surface for band iii, hole pockets of high-mass carriers that cross the Fermi level at the corners of the Brillouin zone (Figure~2e). This feature is clearly represented in the 2D projected cross-sections of the doped Ga:Ge Fermi surface (Figure~S9) and is highly sensitive to the position of the Fermi level. Lowering it by 40~meV to simulate a slight increase in the hole concentration will both enlarge the corner hole pocket projections for band iii and introduce similar features into the projected corners of band ii (Figure~S10).\\

\section*{Epitaxial interfaces and phonon structure}

\begin{figure*}[htbp!]
    \centering
    \includegraphics[width=0.75\linewidth]{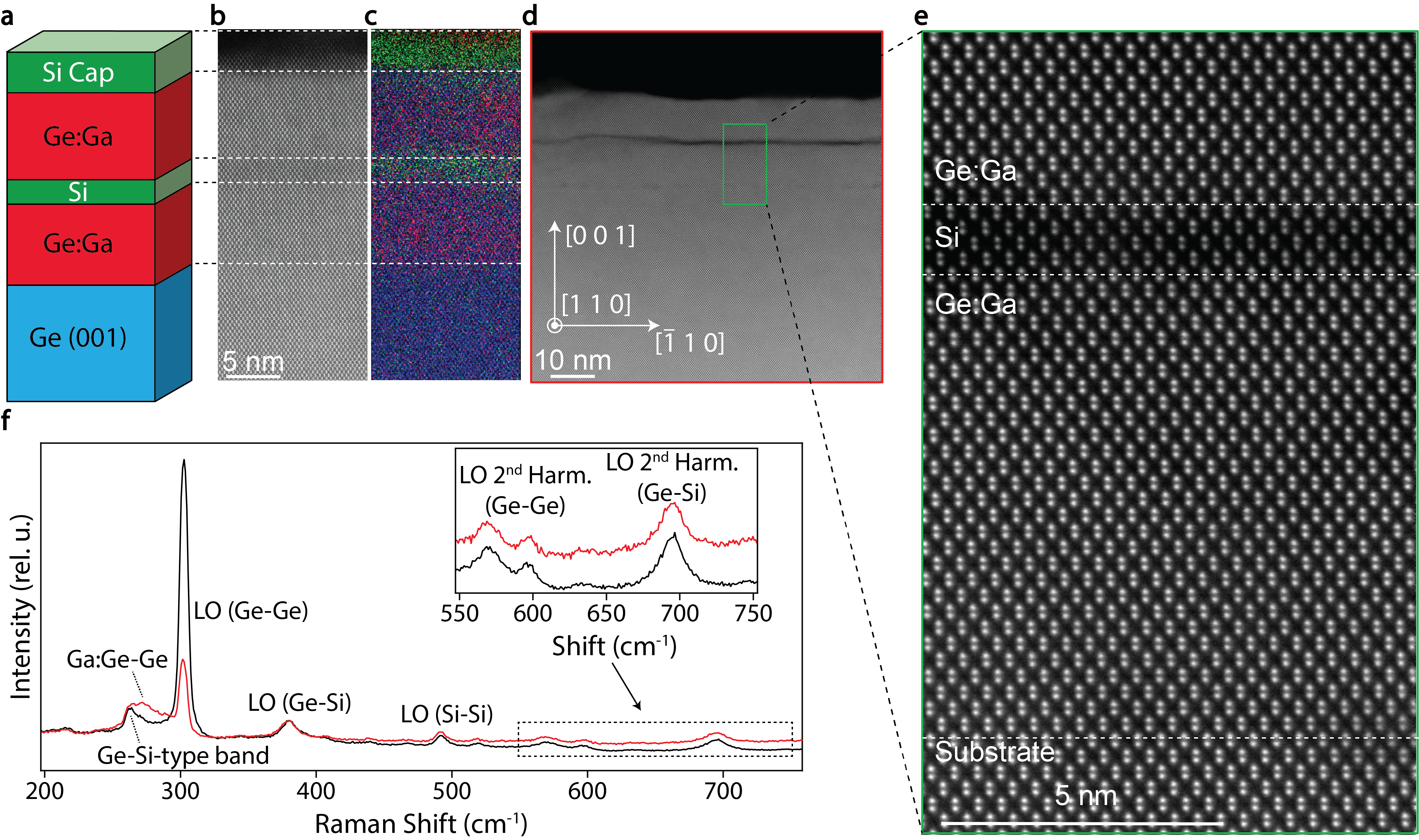}
    \caption{\textbf{Cross-sectional electron microscopy reveals coherent crystalline interfaces.} \textbf{a} Schematic of the film structure as grown by MBE. \textbf{b-c} Cross-sectional TEM/EDS imaging of the $T_\textup{c}$ = 3.5~K hyperdoped sample presenting the compositional profile across the entire film thickness. \textbf{d} Low-magnification cross-sectional HAADF-STEM image of the film. \textbf{e} Zoom-in on the film/substrate and Ga:Ge/Si interfaces, displaying coherent crystalline interfaces. \textbf{f} Raman back-scattering spectra of undoped (black) and doped (red) structures using 532~nm laser excitation at room temperature. The inset highlights the common emergence of highly anharmonic second-order optical bands, identified as the active LO mode here due to the Raman selection rules for backscattering from a (001) Ge surface. For more details regarding the Raman spectra, please see Figure~S12.}
    \label{fig3}
\end{figure*}

Figure~3a presents a schematic structure of the grown film with $T_\textup{c}$ = 3.5~K, in which two hyperdoped Ge layers are grown surrounding a thin Si buffer layer. Cross-sectional transmission electron microscopy (TEM) and energy dispersive spectroscopy (EDS) measurements are presented in Figure~3b-c and show the spatially homogeneous elemental profile of the Ga dopants throughout the hyperdoped layer. Notably, no distinct signatures of Ga clustering are observed compared to earlier reports where annealing was used to activate dopants~\cite{sardashti2021}. Likewise, these MBE-grown materials also exhibit abrupt interfaces and a smooth surface amenable to future planar device fabrication as opposed to previous MBE studies~\cite{strohbeen2023superge}. A view of the atomic structure using low-magnification high-angle annular dark field scanning transmission electron microscopy (HAADF-STEM) is presented in Figure~3d, with a corresponding expansion of the key film/substrate and Ga:Ge/Si interfaces contained in Figure~3e. To verify domain structure, a wider view of the STEM lamella is presented in Figure~S11 where we see no distinct signs of polycrystalline domain formation nor significant surface roughness. Owing to the small amount of mismatch strain and thin layer thicknesses, the growth of Ga:Ge is commensurate with the underlying Ge substrate and the atomic planes on both sides of the interface are in perfect registry. We note that in the current study, we measure a barrier thickness of $\sim$1~nm whereas we estimate a thickness of 5~nm of Si or 10~nm of Ge to be necessary for the demonstration of Josephson tunneling. However, this demonstration of fine thickness control, commensurate interfaces, and homogeneous dopant distribution exhibits the necessary structural and chemical requirements to reduce the disorder in superconductive quantum devices and thus positions these hyperdoped trilayer structures to better mitigate microwave photon loss in quantum information applications~\cite{shim2014bottomup}.\\

We observe that the heavy p-type nature produces a new, softened mode near 280~cm$^{-1}$~\cite{compaan1983raman} in the (zone center) phonon spectra as seen in Raman backscattering spectroscopy (Figure~3f) compared to the undoped case. Details regarding this measurement are presented in the Methods and in Figure~S12. This mode screens the Ge-Ge optical mode, now coupled to the large hole concentration, leading to significant reduction in intensity. Lattice dynamics calculations (Figure~S13) of pure and hyperdoped systems support this observation and reveal phonon softening in hyperdoped Ge. Evaluating the Raman spectra in Figure~3f, both doped and undoped samples yield comparable Raman signals related to the Si-Si and Si-Ge-type sublattices. For the Si-Si optical vibration originating from the thin Si layers, the low dimensionality and diffusion of Ge and Ga atoms into this region lead to a significant mode softening and a strong Si-Ge band near 400~cm$^{-1}$ (Figure~3c)~\cite{fournier-lupien2013}. As both doped and control films exhibit similar Si-Si, Si-Ge, and Ge-Ge bands, we infer high Ga doping does not appreciably affect the growth process. We further compute the Eliashberg spectral function~\cite{Giustino2017} (see Figure~S14) yielding the total coupling strength, $\lambda=0.41$, from which we calculate $T_\textup{c} = 0.77$~K~\cite{AllenDynes1975, McMillan1968}, providing evidence of a conventional phonon-mediated BCS-type pairing mechanism. This coupling primarily originates from the optical phonon modes similar to previous observations in silicon and diamond~\cite{ekimov2004bcsuper, bustarret2006sib}. More details of this calculation are reported in the Methods section.\\

\section*{Dopant-driven crystalline distortion}

\begin{figure*}[htbp!]
    \centering
    \includegraphics[width=0.75\linewidth]{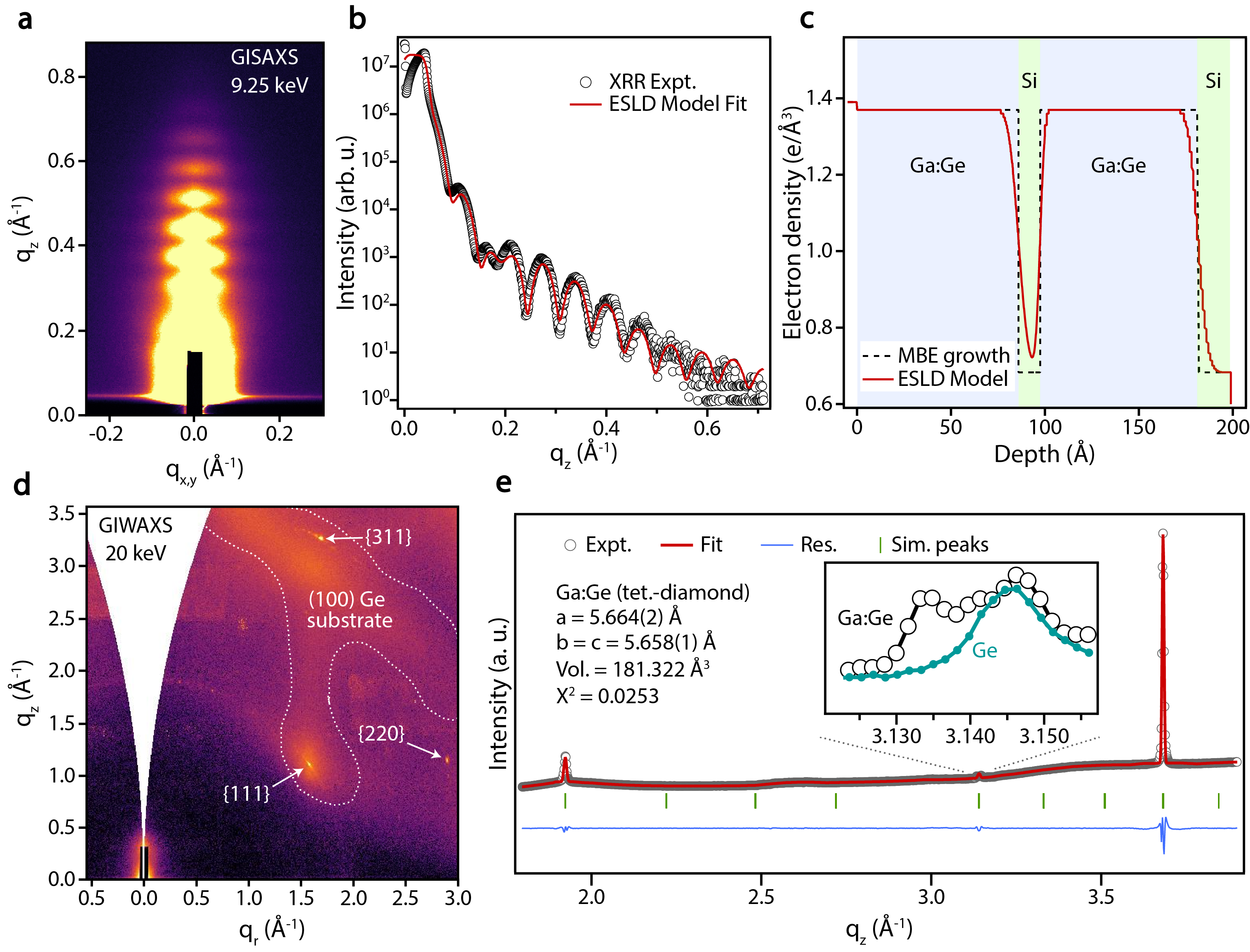}
    \caption{\textbf{X-ray scattering measurement of tetragonal crystalline distortion.} \textbf{a} 2D GISAXS pattern of the Ga:Ge epitaxial film recorded at the critical angle (0.26$^{\circ}$) of the superconducting layer. The intensity map is on a linear scale. \textbf{b} The out-of-plane X-ray reflectivity measurement of the Ga:Ge film with a fit made using the REFLEX software package~\cite{vignaud_reflex_2019}. \textbf{c} Cross-section of the electron scattering length density (ESLD) of the epitaxial stack derived for the fitting model of the fringes in \textbf{b}. \textbf{d} 2D GIWAXS pattern of the Ga:Ge thin film record at the critical angle (0.125$^{\circ}$). The color map is on a log scale. The family of crystallographic planes giving rise to the intense Bragg reflections of the hyperdoped layer are identified with arrows. \textbf{e} Corresponding integrated GIWAXS profile (q$_{xyz}$) and its subsequent structural refinement (Le Bail method). The refined unit cell parameters are displayed and an expansion of the split {220} family of Bragg peaks is shown in the inset to highlight the reduced symmetry of the clamped Ga:Ge epitaxial layer in comparison to the undoped Ge control.}
    \label{fig4}
\end{figure*}

To resolve the microstructure of the hyperdoped heterostructure we use a combination of synchrotron-based grazing incidence X-ray scattering experiments. The 2D grazing-incidence small-angle x-ray scattering (GISAXS) pattern presented in Figure~4a is recorded at the critical angle of the Ga:Ge layers (0.26$^{\circ}$), well above the critical angle of the less dense Si layers (Figure~S15). Kiessig fringes due to the interference of scattered photons within the Ga:Ge film provide evidence of sharp interfaces between the layers. X-ray reflectometry (XRR) analysis is used to derive the electron density profile and compositional grading parameters (rms roughness) of the film in the normal direction. Based on the XRR fit shown in Figure~4b, compositional grading is confirmed between the layers and diffusion is revealed to be asymmetrical with respect to the stack, with larger roughness values determined for Si grown on top of Ga:Ge, compared to Ga:Ge grown on the Si layer. From the XRR fitting, we observe finite diffusion of Ge into the barrier layer, reducing the epitaxial mismatch strain and helping to preserve epitaxial coherence across this interface. Comparing the XRR electron density profile to the characteristic superconducting length scales as calculated (details in the Supplemental information) from the magnetic field dependance of this sample, presented in Figure~S16, we note that the width of the diffusion profile falls well below one coherence length, $\xi$. This is in contrast to the secondary ion-mass spectroscopy (SIMS) profiles in Figure~S17, which suggests a long tail of Ga diffused into the Ge substrate. We believe this disagreement to be the result of the sputtering process smearing out the backside interface during our SIMS measurements. Regardless, in the worst case scenario the Ga:Ge/substrate interface is approximately 1 $\xi$ away from the level at which the Ga dopants fall to a concentration below $1\times 10^{17}$ cm$^{-3}$ (instrumental resolution). Thus, following Anderson's theorem~\cite{anderson1959} we do not expect significant perturbation of the observed superconducting state due to impurity diffusion since the characteristic length scale of our superconductor, $\xi$, is larger than the characteristic diffusion length scale observed in our experiments.\\

The quasi-homoepitaxial growth of the hyperdoped film forces in-plane lattice matching to the substrate, but the larger radius of Ga atoms leads to a predicted out-of-plane tetragonal distortion with the lattice parameter expanding by $\sim$ 0.12\% relative to pure Ge. To experimentally probe this distortion in our system, we employed grazing-incidence wide-angle x-ray scattering (GIWAXS) recorded at the critical angle of Ga:Ge (Figure~S18) to maximize scattering intensity within the hyperdoped layer and take advantage of waveguide-like effects~\cite{steele_how_2023}; Figure~4d. Due to the similar densities of the Ge substrate (5.32~g$\cdot\,$cm$^{-3}$) and the Ga:Ge layer (5.17~g$\cdot\,$cm$^{-3}$), a background signal emerges from the (001)-oriented Ge substrate and is outlined by the dotted lines in the 2D GIWAXS image. Notably, these are only resolved on a log scale and constitute the smoothly changing background of the integrated 1D scattering profile shown in Figure~4e. However, near the critical grazing angle, we observe intense Bragg peaks emerging from the hyperdoped layer, with their indexed peaks highlighted in Figure~4d. As predicted, the symmetry of the hyperdoped layer is reduced due to epitaxial clamping and biaxial stress, as evidenced by the splitting of the \{220\} family of Bragg peaks in Figure~4e, and an asymmetry in the \{311\} peak. Due to the substrate clamping effect, the hyperdoped Ge film adopts a body-centered tetragonal unit cell, similar to the archetypal $\beta$-Sn (tetragonal space group I41/amd). We confirm this asymmetry is not a result of the strained Si layer by conducting similar measurements on the same structure that contains no Ga doping, presented in Figure~S19. Assuming the elastic constants of the hyperdoped layer ($\sim$ 11.22\% Ga substitution) are comparable to pure Ge~\cite{mcskimin1963}, the in-plane lattice mismatch is accommodated by elastic strain and a biaxial stress of $\sigma$ = 106~MPa (see Supplemental Information for details on this calculation).\\

\section*{Conclusions}

Our results establish that superconductivity in MBE-grown Ga:Ge thin films arises from substitutional Ga incorporation rather than Ga-clustering, allowing for the creation of coherent superconductor-semiconductor interfaces. TEM and EDS imaging confirm pristine structural and compositional homogeneity. Synchrotron-based methods confirm Ga occupies substitutional sites within the Ge lattice and the interplay between Ga incorporation and epitaxial constraints (substrate clamping) leads to a subtle tetragonal ($\beta$-Sn) distortion, in line with our DFT results. Electronic structure calculations show that Ga incorporation at substitutional sites shifts the Fermi level deep into the valence band and flattens the electronic bands near the R point, likely promoting superconductivity. These results show significant promise for future device implementation of superconducting group IV materials; however, challenges yet remain. The band offsets of these superconducting materials interfaced with their semiconducting counterparts and the effect of long-range disorder (large kinetic inductance) in device applications are unable to be fully addressed in the current study. Looking ahead, it is necessary to further develop the proof-of-principle trilayer structures discussed here into full-fledged Josephson junction devices. Furthermore, an equivalent Ga floating layer forms during silicon doping~\cite{kesan_dopant_1991} allowing for a natural extension of this study towards growth of superconducting silicon. This would enable the trade of relatively lossy germanium substrates for state-of-the-art low loss silicon substrates for use in quantum information devices. The findings presented here highlight the intrinsic nature of superconductivity in hyperdoped Ga:Ge thin films and underscore the potential of group IV covalent superconductors as a scalable platform for epitaxial superconductor-semiconductor heterostructures.\\


\section*{Acknowledgments}
S.S.-R. acknowledges Dan Huber for assisting with TEM sample preparation. P.J. acknowledges fruitful discussions with Ross McKenzie and Ben Powell. P.J.S and J.S. acknowledge funding support from the United States Air Force Office of Scientific Research award number FA9550-21-1-0338. We acknowledge computational resources at the Australian National Computational Infrastructure and Pawsey Supercomputing Research Centre through the National Computational Merit Allocation Scheme. This work was partially supported by the Australian Research Council under the following grants: DE230100173 (J.A.S.), LP210200636 (Y.-H.C., P.J.), DE220101147 (C.V.). This research was undertaken on the X-ray Absorption Spectroscopy and the SAXS/WAXS beamlines at the Australian Synchrotron, part of ANSTO. The XRR data reported in this paper were obtained at the Central Analytical Research Facility operated by Research Infrastructure (QUT). Electron microscopy was performed at the Center for Electron Microscopy and Analysis (CEMAS) at The Ohio State University.
 
\section*{Author contributions:}
Conceptualization: P.J., J.S., and E.D. Methodology: J.A.S., P.J.S., C.V., S.S.-R. Investigation: J.A.S., P.J.S., C.V., A.B., A.D., Y.-H.C., J.V.D., F.K., A.L., D.P., S.S.-R. Visualization: J.A.S., P.J.S. Funding acquisition: J.A.S., P.J., J.S. Project administration: J.A.S., P.J.S., P.J., J.S. Supervision: J.A.S., P.J., J.S., E.D., L.W. Writing – original draft: P.J., J.A.S., P.J.S. Writing – review and editing: J.A.S., P.J.S., C.V., P.J., J.S.

\section*{Competing interests:}
There are no competing interests to declare.
\clearpage


\begin{widetext}
\section*{Methods}

\textbf{Thin film growth.} The materials used in this study are all grown in a custom Varian Gen II molecular beam epitaxy (MBE) system on 50.8~mm undoped Ge (001) wafers. Prior to loading into vacuum, the wafers are etched in deionized water at 90~$^{\circ}$C for 15 minutes and then immediately loaded into the vacuum chamber on indium-free mounting blocks. The wafers are sequentially outgassed inside the growth reactor at 250~$^{\circ}$C, 450~$^{\circ}$C, and finally 650~$^{\circ}$C, for 15~minutes, 15~minutes, and~5 minutes, respectively, then cooled to room temperature for growth. Pure germanium (6N) and silicon (3N) source material is deposited via Thermionics HM2 e-Guns operating at a 6~kV acceleration voltage. Gallium doping is done with a standard Knudsen effusion cell source (MBE Komponenten). Atomic fluxes are measured prior to growth with a retractable quartz crystal microbalance (QCM) placed in the center of the growth path in front of the substrate. The substrate is faced away from the atomic beam path(s) during flux measurement. Reported temperatures are from a thermocouple attached close to the backside of the substrate that has been calibrated to the GaAs (001) oxide-removal temperature. Substrate temperatures reported in Figure~S3 are reported such that at temperatures lower than 95$^{\circ}C$, there is no active heating during film growth. The thermal mass of the manipulator maintains the desired temperature within 5$^{\circ}C$ throughout the entire film growth $\sim$ 10-20 minutes. Samples grown at 95$^{\circ}C$ and above have their substrate temperatures controlled via PID temperature controller. Samples are grown with a Ge flux of $\sim 8 - 8.5 \times 10^{13}~cm^{-2},s^{-1}$, a Ga flux of $1.4 - 1.9 \times 10^{13}~cm^{-2},s^{-1}$, and a Si flux of $2.5 - 3 \times 10^{13}~cm^{-2},s^{-1}$ as measured by our QCM. The nominally 20~nm thick films are grown as follows, as described in Ref. \cite{strohbeen2023superge} : (1) 10~nm of Ga-doped Ge, (2) 0.5~nm of Si, (3) 10~nm of Ga-doped Ge, (4) 1~nm Si cap. For samples grown where the barrier layer and capping layer is made of germanium, the barrier and cap thicknesses are 10~nm and 5~nm, respectively. 

\textbf{Hall bar device fabrication and measurement.} A 5$\times$5 chip is first cleaned in acetone for 2~min, followed by isopropyl alcohol for 1~min, and then blow dried with compressed air. The chip is then dehydrated on a hotplate in a fume hood at 110$^{\circ}$C for 1 minute and then immediately placed into a spin coater. AZ 5214 photoresist (PR) is spun onto the chip at 4000 rpm for 60~s. Edge beads are removed using acetone and then the chip is placed back onto the hotplate at 110$^{\circ}$C for 1~min for the PR post-bake. The photolithography was conducted with a S\"{u}ss MJB3 mask aligner using hard contact. The PR was exposed for 9~s, developed for 30~s in AZ 1:1 developer, and then rinsed in DI water for 15~s. The Hall bar was etched in an Oxford PlasmaPro System 100 Cobra inductively-coupled plasma (ICP) dry etcher system using a chlorine gas chemistry of 5 sccm Cl$_{2}$, 5 sccm BCl$_{3}$, 5 sccm Ar for 30~s at an ICP power of 500~W and a 75~W bias. This etch results in a step height of 140~nm as measured in an atomic force microscope, confirming a deep etch into the Ge substrate. The resist is stripped in AZ 400T PR stripper on a hotplate at 80$^{\circ}$C for 20 minutes, cleaned in isopropyl alcohol, and then blow dried. The device is wire bonded to a custom daughterboard and then cooled down in an Oxford Triton dilution refrigerator with a base temperature of 15~mK and a 14~T single-axis magnet.\\

\textbf{Four-point transport measurements.} Electrical characterization is done in an Oxford TeslatronPT He4 cryostat using a HelioxVT He3 probe insert with a base temperature of $\sim$ 270~mK (HelioxVT) and magnetic field capabilities up to 12~T. Measurements are collected in a similar fashion to Ref.~\cite{strohbeen2023superge}, using a standard Van der Pauw wiring configuration on square pieces cleaved from near the center of each wafer. On-chip contacts are made via annealed In-Sn eutectic alloy at each of the four corners which are then wired to a daughterboard using gold wire.\\

\textbf{First-principles calculations.} First-principles calculations were performed using DFT as implemented in VASP \cite{kresse1993abinitmd, kresse1996abinitpw}, except for the electron-phonon coupling calculations where we employed Quantum ESPRESSO~\cite{QE2017}. In VASP, projector augmented wave (PAW) pseudopotentials including semicore 3d electrons were adopted, with a plane-wave cutoff of 550~eV. A $7 \times 7 \times 7$ $\Gamma$-centered k-point mesh was used to sample the Brillouin zone of the conventional 8-atom unit cell of Ge (cubic diamond structure, space group $Fd\overline{3}m$). Structural optimizations were performed using the PBEsol exchange-correlation functional~\cite{perdew2008}, which yields lattice parameters in excellent agreement with the experiment (see Table~S1), and a convergence threshold of 0.005~eV/\r{A} on all forces. 

For the Ga-doped Ge structure with substitutional fine-structure, one host atom in the 8-atom unit cell is replaced by a Ga atom, corresponding to a doping concentration of 12.5\%. Conversely, interstitial doping is modeled by placing one Ga atom at interstitial doping sites, representing a doping concentration of 11.1\%. To model the effect of substrate clamping in Ga-doped Ge, the in-plane lattice parameters were fixed to the pristine Ge values, and only the out-of-plane lattice parameter was allowed to relax. K-edge spectra were simulated using the supercell core-hole method~\cite{karsai2018eph} within a $2\times2\times2$ supercell containing 64 atoms (72 atoms in the case of interstitial doping). The formation energy of a neutral Ga dopant in the Ge crystal is calculated as $E_{\textup{Form}} = E_{\textup{d}} - E_{\textup{p}} - \sum_{i}n_{i}\mu_{i}$, where $E_{\textup{d}}$ is the total energy of the structure containing the defect, $E_{\textup{p}}$ is the total energy of the pristine structure, $n_{i}$ is the number of atoms of species $i$ being added ($n>0$) or removed ($n<0$), and $\mu_{i}$ is the chemical potential. For Ga, this is the energy per atom of a Ga crystal (base-centered orthorhombic structure).

To calculate the electronic band structure and density of states, we adopted the MBJLDA meta-GGA functional~\cite{tran2009smbg} based on the modified Becke-Johnson exchange functional. This functional has been shown to yield excellent results for the band structure and band gap of many semiconductors, including Ge~\cite{rodl2019hexge, fadaly2020hexIVemission}, which standard semilocal DFT predicts erroneously to be metallic. Spin-orbit coupling was included. A denser $20\times20\times20$ $k$-mesh was employed to compute the density of states. To accurately determine the Fermi-level shifts in Ga-doped Ge, a $31\times31\times31$ $k$-mesh was used. Vibrational properties (shown in Fig.~S15) were computed using 
finite differences~\cite{Togo2015} within supercells containing 216 atoms 
($3\times3\times3$ supercells of the 8-atom unit cell), employing $3\times3\times3$ 
$k$-point meshes and displacement amplitudes of 0.015~\AA.

We used Quantum ESPRESSO~\cite{QE2017} to perform the electron-phonon coupling calculations, using norm-conserving pseudopotentials from the PseudoDojo library v0.4.1~\cite{vanSetten2018} with standard accuracy and the PBEsol exchange-correlation functional, and a plane-wave cutoff of 80~Ry (1088~eV). We adopted the same model of substitutionally Ga-doped Ge, ignoring for simplicity the effect of substrate clamping. Relaxing the unit cell after Ga substitution results in a small lattice expansion of 0.09\%. Spin-orbit coupling was neglected. We verified that the resulting Fermi level shift into the valence bands (0.96~eV) is consistent with the one obtained with the MBJLDA functional within VASP when spin-orbit coupling is neglected (0.93~eV).
We calculated the Eliashberg spectral function $\alpha^2F(\omega)$ defined as~\cite{Giustino2017}:
\begin{equation} \alpha^2F(\omega)=\frac{1}{N_\textup{F}N_\mathbf{k}N_\mathbf{q}}\sum_{mn\nu,\mathbf k\mathbf q} |g_{mn\nu}(\mathbf k,\mathbf q)|^2 \,\delta(\varepsilon_{n\mathbf k}-\varepsilon_\textup{F}) \delta(\varepsilon_{m\mathbf k+\mathbf q}-\varepsilon_\textup{F}) \delta(\hbar\omega-\hbar\omega_{\mathbf q\nu}), \label{eq:a2F}
\end{equation}
where $N_\textup{F}$ is the density of states at the Fermi level, $N_\mathbf{k}$ and $N_\mathbf{q}$ are the total number of $\mathbf k$ and $\mathbf q$ points used to sample the Brillouin zone, $\varepsilon_{n\mathbf k}$ is the electron energy with band index $n$, $\varepsilon_\textup{F}$ is the Fermi energy, and $\hbar\omega_{\mathbf q\nu}$ is a phonon energy with momentum $\mathbf q$ and branch $\nu$. $g_{mn\nu}(\mathbf k,\mathbf q)$ are the electron-phonon matrix elements~\cite{Giustino2017} calculated using density-functional perturbation theory.
The total electron-phonon coupling strength $\lambda$ is then obtained as $ \lambda=2\int d\omega\frac{\alpha^2F(\omega)}{\omega}$. 
Using the semi-empirical Allen-Dynes-McMillan equation, the superconducting critical temperature $T_\textup{c}$ is estimated as~\cite{McMillan1968,AllenDynes1975}:
\begin{equation} 
k_\textup{B}T_\textup{c}=\frac{\hbar\omega_\textup{log}}{1.2}\,\mbox{exp}\left[\frac{-1.04(1+\lambda)}{\lambda-\mu^\ast(1+0.62\lambda)}\right], \label{eq:Tc}
\end{equation}
where $k_\textup{B}$ is Boltzmann's constant, $\omega_{\textup{log}}$ is the logarithmic average of the phonon frequencies, given by 
$\omega_{\textup{log}}=\mbox{exp}\left[\frac{2}{\lambda}\int_0^\infty d\omega\frac{\alpha^2F(\omega)}{\omega}\log\omega\right]\!,$
and $\mu^\ast$ is the Coulomb pseudopotential, here set to 0.1~\cite{AllenDynes1975}. To converge the Brillouin-zone integrals in 
Eq.~\eqref{eq:a2F}, we employed a $20\times20\times20$ and $5\times5\times5$ $k$ and $q$-mesh, respectively, and a Gaussian broadening of 0.04~Ry for the electronic double-delta functions.\\

\textbf{Scanning transmission electron microscopy.} Cross-sectional specimens for scanning transmission electron microscopy (S/TEM) analysis were prepared using a focused ion beam (FIB) system (FEI Helios NanoLab 600 DualBeam). To minimize surface damage, the samples were thinned and polished with Ga ions at 5~kV. S/TEM imaging and energy-dispersive X-ray spectroscopy (EDS) were conducted using a Thermo Fisher Scientific Themis-Z microscope. The instrument operated at an accelerating voltage of 300~kV with a semi-convergence angle of 20~mrad. HAADF-STEM images were collected using a 64-200~mrad high-angle annular dark field (HAADF) detector. High-resolution images were acquired through multiple rapid scans (2048 $\times$ 2048~pixels, 200~ns~per~pixel), which were combined to improve the signal-to-noise ratio. EDS analysis was performed with the Super-X EDS detector, and the elemental mappings are presented as net count images.\\

\textbf{Raman Spectroscopy.} Raman spectra were recorded from the thin film surface in a backscattering configuration, using an integrated Edinburgh confocal micro-Raman instrument (RM5). Excitation was provided by one of three solid-state laser lines (532~nm, 638~nm and 785~nm) and focused on the surface using a long working lens (Olympus, 10x, 0.4~NA). Dispersion was achieved through a 1200 g/mm diffraction grating (spectral resolution 0.2~cm$^{-1}$), instrument calibration was verified through checking the position of the Si band at $\pm$520.7~cm$^{-1}$, and spectra were recorded using a thermoelectric CCD detector. Changes to the local vibrational structure of the hyperdoped film are first confirmed through the evaluation of the excitation energy dependence of the Raman back-scattering signals (Figure~S12). At longer excitation wavelengths (785~nm), both the control and target films exhibit comparable Raman response, while more surface-sensitive information is recovered at shorter wavelengths (532~nm) to resolve differences in the thin films (Figure~3f).\\

\textbf{Synchrotron-based GISAXS/GIWAXS.} To identify the crystalline phases and microstructure of films, synchrotron-based grazing incidence wide-angle X-ray scattering (GIWAXS) data were collected at the small angle X-ray scattering/wide angle X-ray scattering (SAXS/WAXS) beamline at the Australian Synchrotron~\cite{kirby2013xrayscatter}: 2D scattering patterns were recorded using a wavelength of either 1.340375~\r{A} (9.25~keV) or 0.619924~\r{A} (20~keV), using an energy-selective Pilatus 2M detector. The measurements made using a 20~keV beam energy were to increase the observed Q-range for the structural refinements, using an energy threshold of 15~keV to suppress the X-ray fluorescence signals arising from the excited K-edges of Ge (11.1~keV) and Ga (10.37~keV). Experiments at 9.25~keV had no such issues. Two different sample-to-detector distances were employed to record both SAXS (5.5~m) and WAXS (0.45~m) profiles, each calibrated using a silver behenate reference standard. The sample and detector were enclosed in a vacuum chamber to suppress air scatter. Scattering patterns were measured as a function of the angle of incidence, with data shown acquired with an angle of incidence near the critical angle (for a given X-ray energy) to maximize scattering intensity from the sample. All collected 2D images were azimuthally integrated using PyFAI~\cite{ashiotis2015pyfai} and processed using a custom Python routine. The resulting unit cell models were refined using the La Bail method implemented in Fullprof~\cite{RODRIGUEZCARVAJAL199355}, a comprehensive analysis software. In line with our other characterization techniques, our GIWAXS experiments failed to detect Ga droplet formation or segregation within the film. Due to the relatively small scattering volume of the Si layers, no additional peaks are indexed to this phase either.\\

\textbf{Synchrotron-based EXAFS experiments.} Ga K-edge X-ray absorption spectra (XAS) were captured at the Australian Synchrotron (ANSTO in Clayton, Victoria) MEX1 beamline in a similar fashion to previously reported experiments~\cite{zhang2024zndopant}. A double-crystal Si (111) monochromator equipped with focusing optics was used to reduce harmonic content while producing excitation energy. An inline Ga$_2$O$_3$ reference (the initial peak of the first derivative occurs at 10.3687 keV) was used to calibrate the monochromator at Ga-K absorption edge with E0 set to its known value of 10.3671~eV. All acquisitions were performed on thin film materials in a fluorescent mode. Brief XAS scans of specific samples were conducted to verify the stability of the materials under at least several minutes of X-ray exposure before the full acquisitions. Samples were measured with varying energy intervals over the pre-edge (5~eV) and the XANES region (0.25), with 0.035 intervals in $k$-space over the EXAFS. At low-$k$ an integration time of 1~s was employed per interval, with longer integration times weighted toward high-$k$ portions of the spectrum, up to a maximum value of 9 in $k$-space (max of 2~s). Multiple scans were collected at different positions of the sample. Data processing including background subtraction, scan averaging, edge-height normalization, and rebinning was performed based on the Athena program~\cite{ravel2005ifeffit}. For normalization of energy spectra and removal of background, the pre-edge range was set between -150~eV and -30~eV, while the normalization range spanned from 80 to 300 eV post-edge. The order of normalization was designated as 3. For the presented EXAFS data, the $k$-weight was configured as 2, and the $k$-range for the forward Fourier transform was defined from 0.8 to 8.5.\\

\textbf{EXAFS fine-structure modeling.} Fits were conducted in the ARTEMIS, part of the IFEFFIT software package~\cite{ravel2005ifeffit} using a similar procedure to previous studies~\cite{zhang2024zndopant}. The scattering paths used to evaluate and model the data were derived from the optimized crystal structures with different defects. Using these input structures to derive the scattering paths, the only agreeable fine structure accounting for the dominant EXAFS signal arising from an R fitting window of 1.8 to 5.1~\r{A} emerged using the calculated Ga substituted structure. To simulate the Ga substituted Ge crystal fine structure, the input paths out to 5~\r{A} are determined using the DFT-derived Ga substitutionally doped Ge structure and the FEFF package~\cite{ravel2005ifeffit}, with a distance fuzz = 0.030~\r{A}. This generated 6 unique scattering pairs centered on the Ga substitutional site, with some other lower ranked paths omitted from the actual model to avoid over-fitting (see Table~S2). A common scaling factor has been used on all path distance corrections to account for the difference between the DFT simulated unit cell volume and the experiment crystal volume at room temperature. A phase correction has been made to Figure~2d in the main text using the most intense Ga-Ge single scattering path, near 2.45~\r{A}. The data were ultimately fit using 18 variables and 26 independent points, resulting in a reasonable R-factor of 0.015. The fitting was conducted in R-space using multiple k-weightings, with k = 1, 2 and 3. Errors of individual fit parameters were determined using ARTEMIS to take into account the correlations between parameters and known parameters without error estimates were fixed during the fit.\\

\textbf{X-ray Reflectometry.} The X-ray reflectometry patterns of the thin film samples were collected using a Rigaku SmartLab X-ray diffractometer with a goniometer radius of 300~mm, under CuK$\alpha$ radiation ($\lambda$ = 1.54059~\r{A}, 40~kV 40~mA). The X-ray was converged into a parallel beam using a parabolic multi-layer mirror in a CBO-PB module, followed by a channel-cut Ge 2x(220) monochromator. A fixed incident slit of 0.1 mm was used for fine angular resolution. The thin films were precisely aligned according to its specular reflection on top of a RxRy double tilt attachment on a $\phi$ rotation stage. A 0.2~mm anti-scattering slit and a 0.3~mm detector slit were used to enhanced the signal intensity. Both the primary and secondary sides use a 5$^{\circ}$ soller slit to control axial divergence. Each pattern was collected by a Hypix3000 detector in 0D mode from 0 to 10~$^{\circ}2\theta$ with a 0.01$^{\circ}$ step size and at a speed of 0.1~$^{\circ}2\theta,\cdot,min^{-1}$. Fits to patterns were made using the REFLECT standalone reflectivity fitting software~\cite{vignaud_reflex_2019}.\\

\newpage

\section*{Supplemental Information}
\beginsupplement
\setcounter{figure}{0}
\setcounter{section}{0}

\section{State of hyperdoping in germanium}

\begin{figure}[htbp!]
    \centering
    \includegraphics[width=0.8\linewidth]{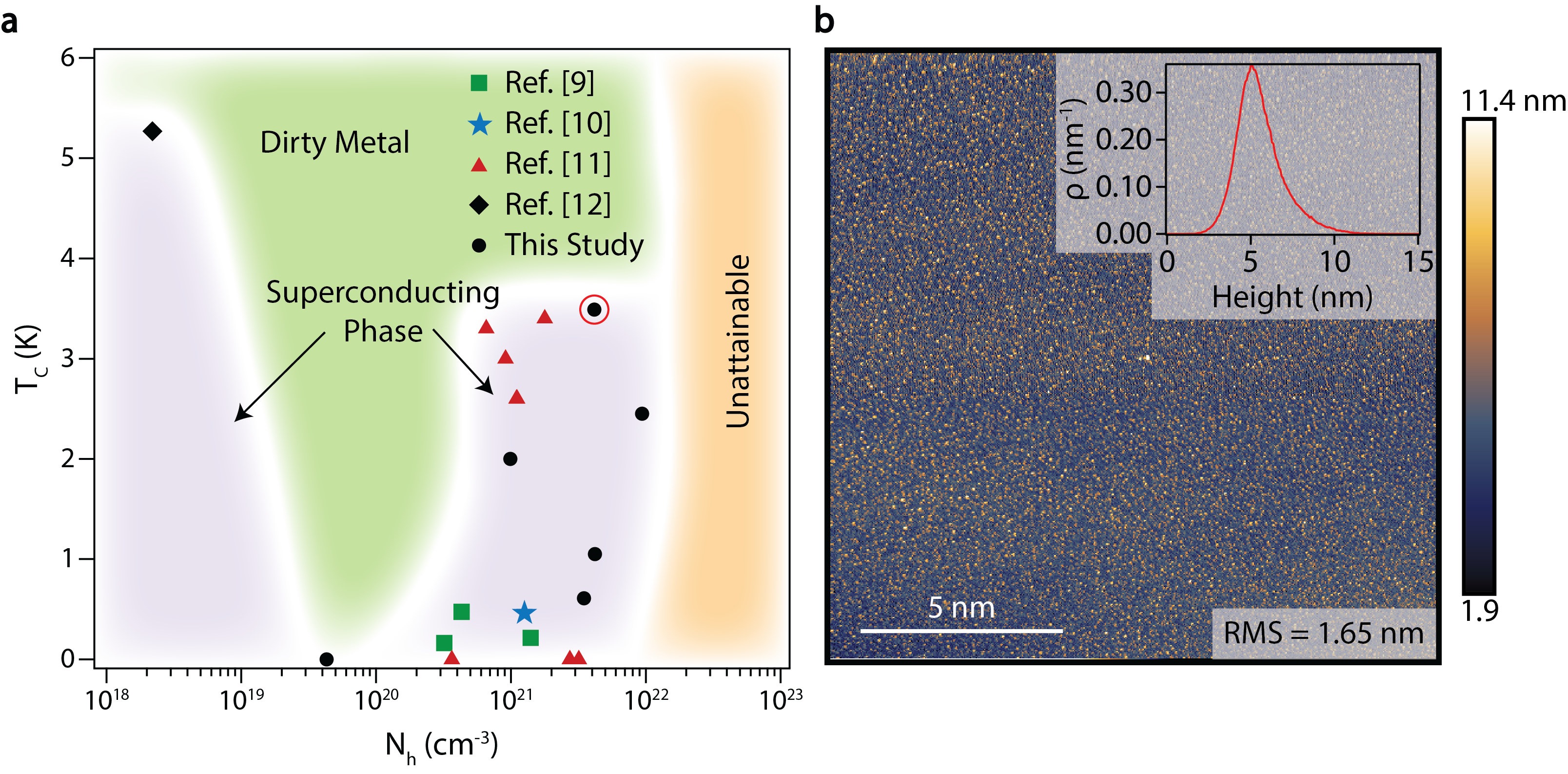}
    \caption{\textbf{a} Comparison of this study against literature reports for superconducting $T_\textup{c}$ as a function of Hall carriers in Ga-doped germanium. The colored star, square, and triangular points were prepared via ion implantation~\cite{herrmannsdorfer2009gage, prucnal2019, sardashti2021} while the black circles (this study) and black diamond~\cite{strohbeen2023superge} are MBE-grown samples. \textbf{b} Representative AFM image of the surface of the sample circled in red in \textbf{a}. We observe an RMS roughness of roughly 1.65~nm, a significant improvement over previous MBE reports~\cite{strohbeen2023superge}. The inset presents the normalized height distribution, $\rho$, across the entire 15~$\mu m \times$15~$\mu m$ image.}
    \label{S1}
\end{figure}

\clearpage
\section{Material optimization}

\begin{figure}[htbp!]
    \centering
    \includegraphics[width=0.7\linewidth]{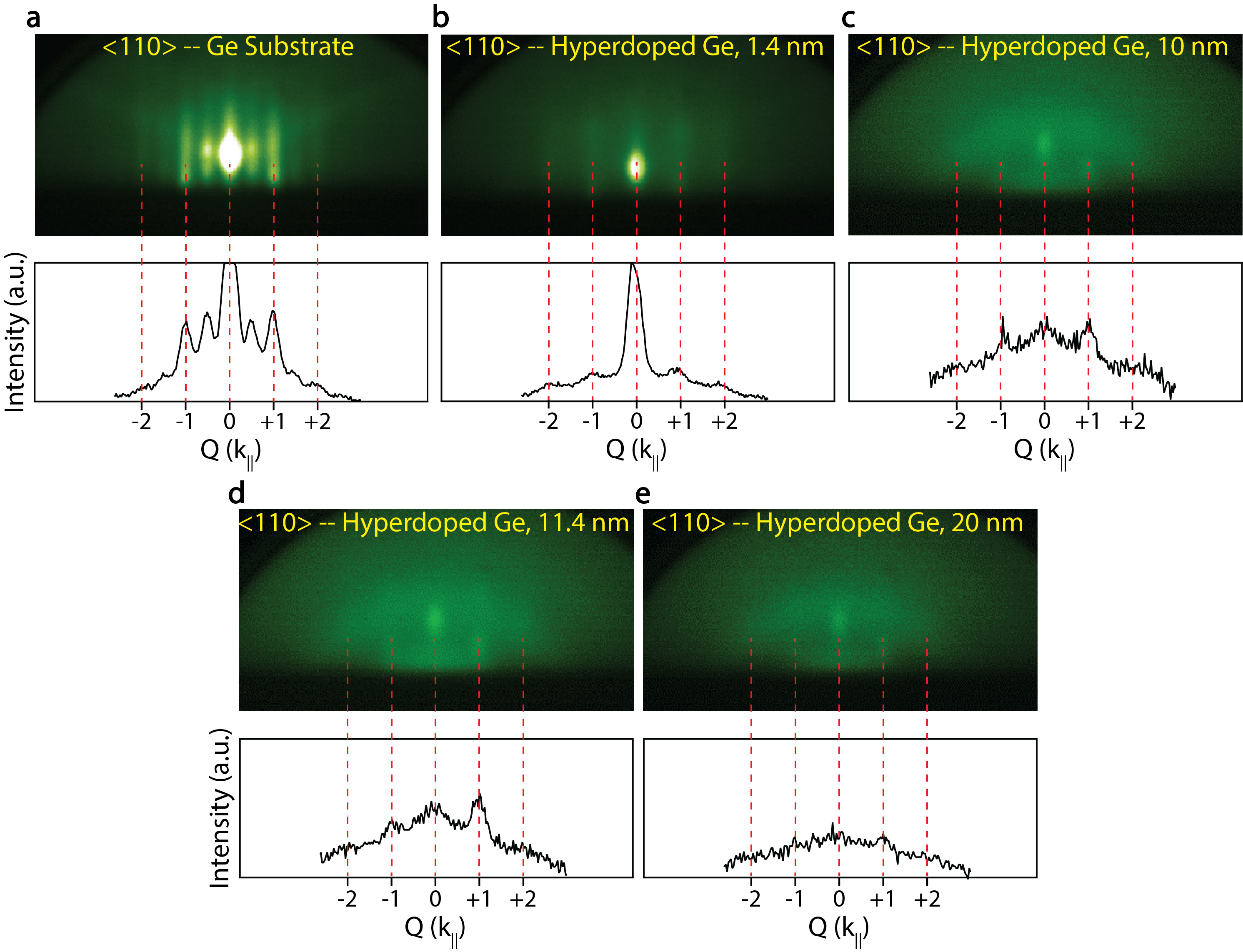}
    \caption{RHEED line cuts for superconducting hyperdoped Ge thin film. \textbf{a} Ge substrate $\langle 110\rangle$-type direction, taken at room temperature just before deposition of Ga:Ge film. \textbf{b} Initial surface of Ga:Ge film. \textbf{c} RHEED pattern after the first 10nm of Ga:Ge deposited. \textbf{d} After the Si spacer layer and initial growth of the second Ga:Ge layer. \textbf{e} End of Ga:Ge growth, before Si cap is deposited. We note a superimposed amorphous ring over a faint $\langle 110\rangle$-type direction diffraction pattern. This is consistent with previous observations that Ga-adsorption at low temperatures creates a floating layer at the surface from which the incorporated dopants are taken~\cite{kesan_dopant_1991}.}
    \label{S2}
\end{figure}

Film growth is monitored using reflection high-energy electron diffraction (RHEED) along a $\langle110\rangle$ cut. Representative RHEED images throughout the growth steps are presented in Figure~\ref{S2}. The haziness observed in Figure~\ref{S2}b-e is caused by a surface floating layer of Ga that forms during the growth process~\cite{kesan_dopant_1991}. The Ga shutter is closed during Si deposition, however, we note that the floating layer slightly incorporates into the Si layer as indicated by the increased intensity in the specular RHEED reflection in Figure~\ref{S2}d compared to Figure~\ref{S2}c. To reduce parasitic heating of the substrate, the hot Ga crucible was backed up within the vacuum chamber by 4~in to reduce radiative heating of the growth surface when the shutter is open.\\

\begin{figure}[htbp!]
    \centering
    \includegraphics[width=0.9\linewidth]{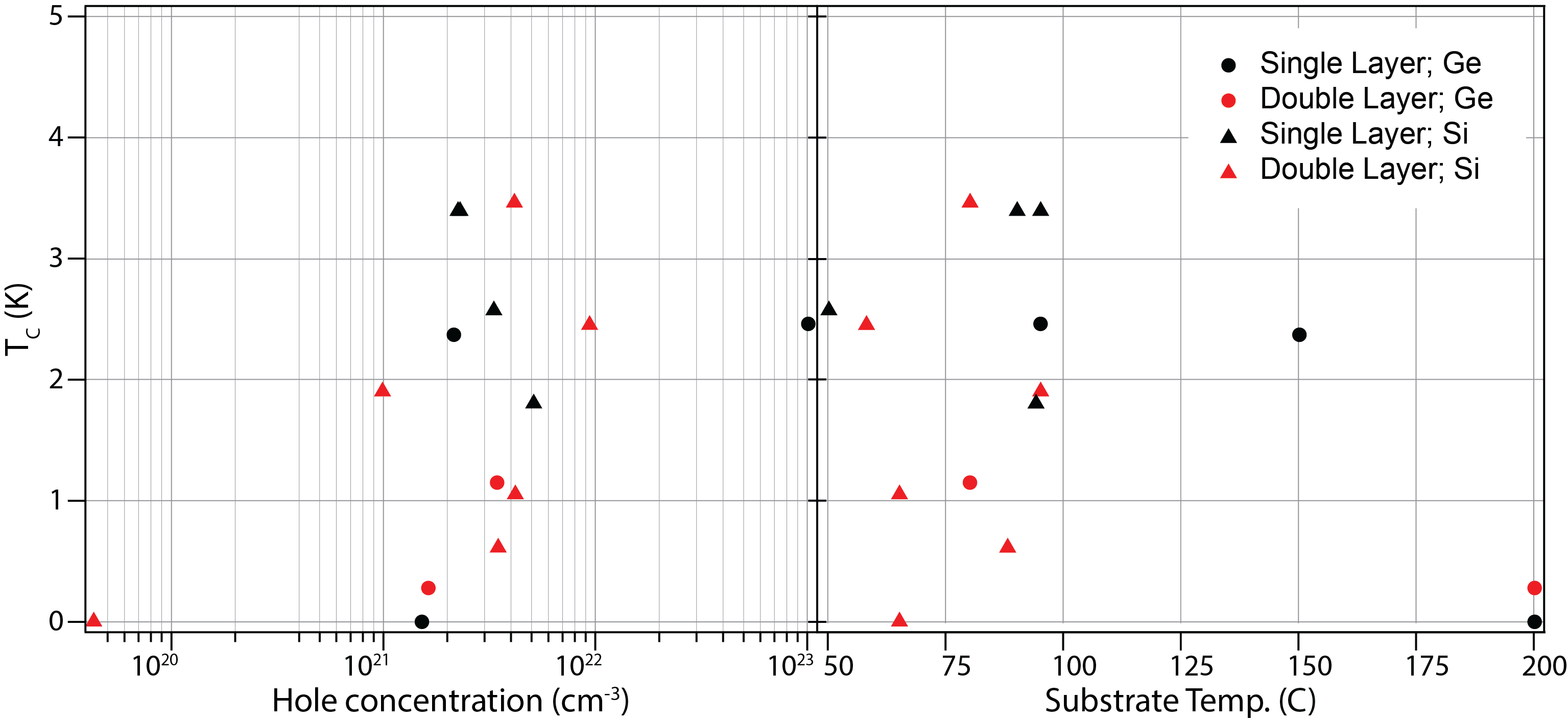}
    \caption{Superconducting $T_\textup{c}$ comparing the effects of silicon as a function of hole concentration and growth temperature. The circular data points are the samples where we replaced the Si tunnel barrier and capping layers with germanium while the triangular points are the samples that contain silicon. The black points are samples in which just a single layer of hyperdoped germanium is grown and the red points are the samples where we grew bilayer heterostructures. We observe a tendency for Si-containing samples to exhibit marginally increased carrier concentration and superconducting $T_\textup{c}$. A trend in $T_\textup{c}$ as a function of growth temperature is further demonstrated where we optimize the superconducting transition temperature at growth temperatures of 95-150$^{\circ}$C.}
    \label{S3}
\end{figure}

The scatter in $T_\textup{c}$ observed in Figure~\ref{S3} for samples grown at similar conditions is attributed to small variations in atomic fluxes caused by drift during the course of film growth as well as the slight turbulence present within the Ge melt~\cite{mclean1998turbulence}. In principle, changing the Ge deposition source from an electron beam deposition source to a different technology would help mitigate these variations.\\

\begin{figure}[htbp!]
    \centering
    \includegraphics[width=0.9\linewidth]{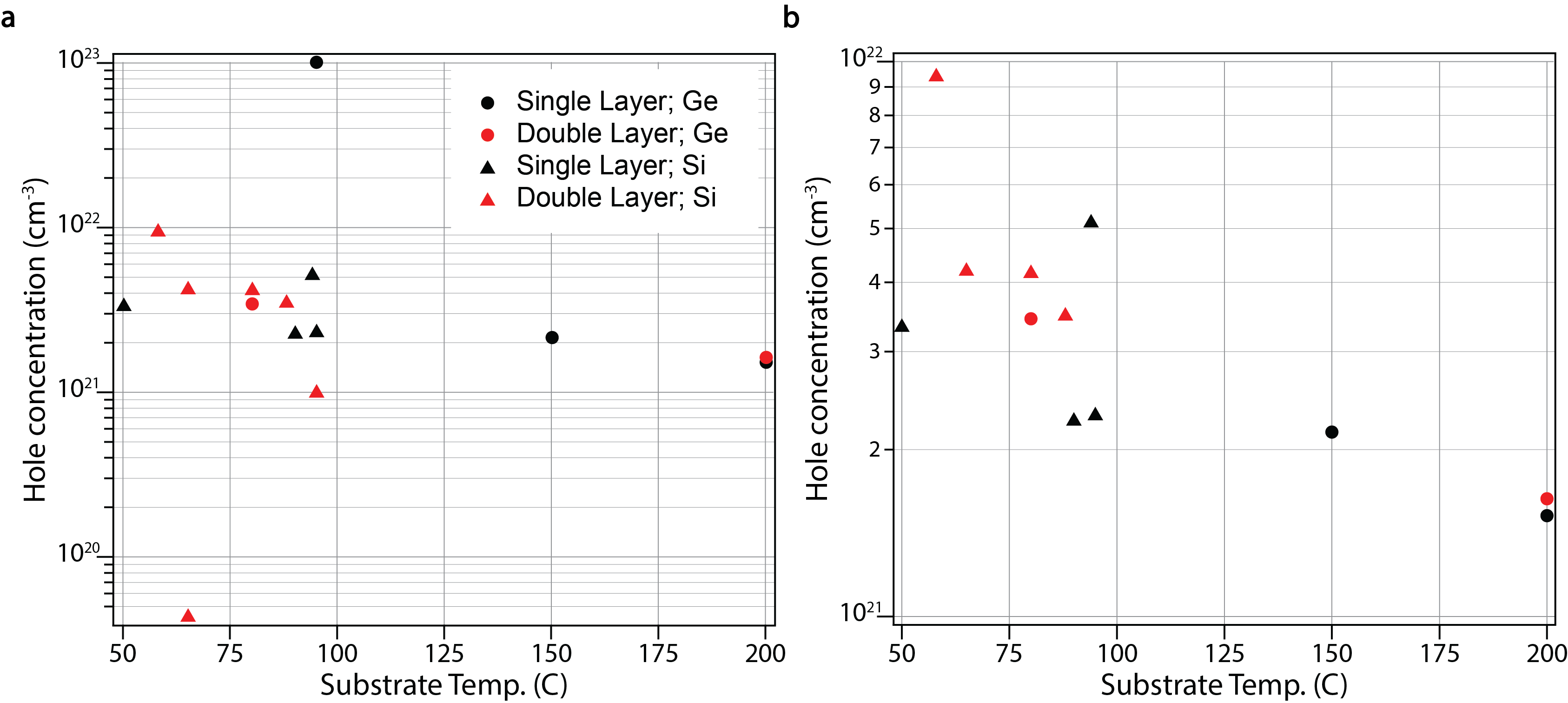}
    \caption{Hole concentration as a function of substrate temperature during growth. The same data point identification in Figure~\ref{S3} is also used here. \textbf{a} Presents the entire set of samples, where we observe two outlier samples: one in which carrier concentration is drastically reduced compared to the other samples and one with significantly higher carrier concentration. We speculate poor dopant activation and Ga droplet formation as the cause of these two samples, respectively. We zoom in on the rest of the samples in \textbf{b}, where we see that a general downward trend is observed as substrate temperature increases. This trend is consistent with an increased growth temperature promoting Ga segregation, reducing the carrier concentration of the films.}
    \label{S4}
\end{figure}

\begin{figure}[htbp!]
    \centering
    \includegraphics[width=0.9\linewidth]{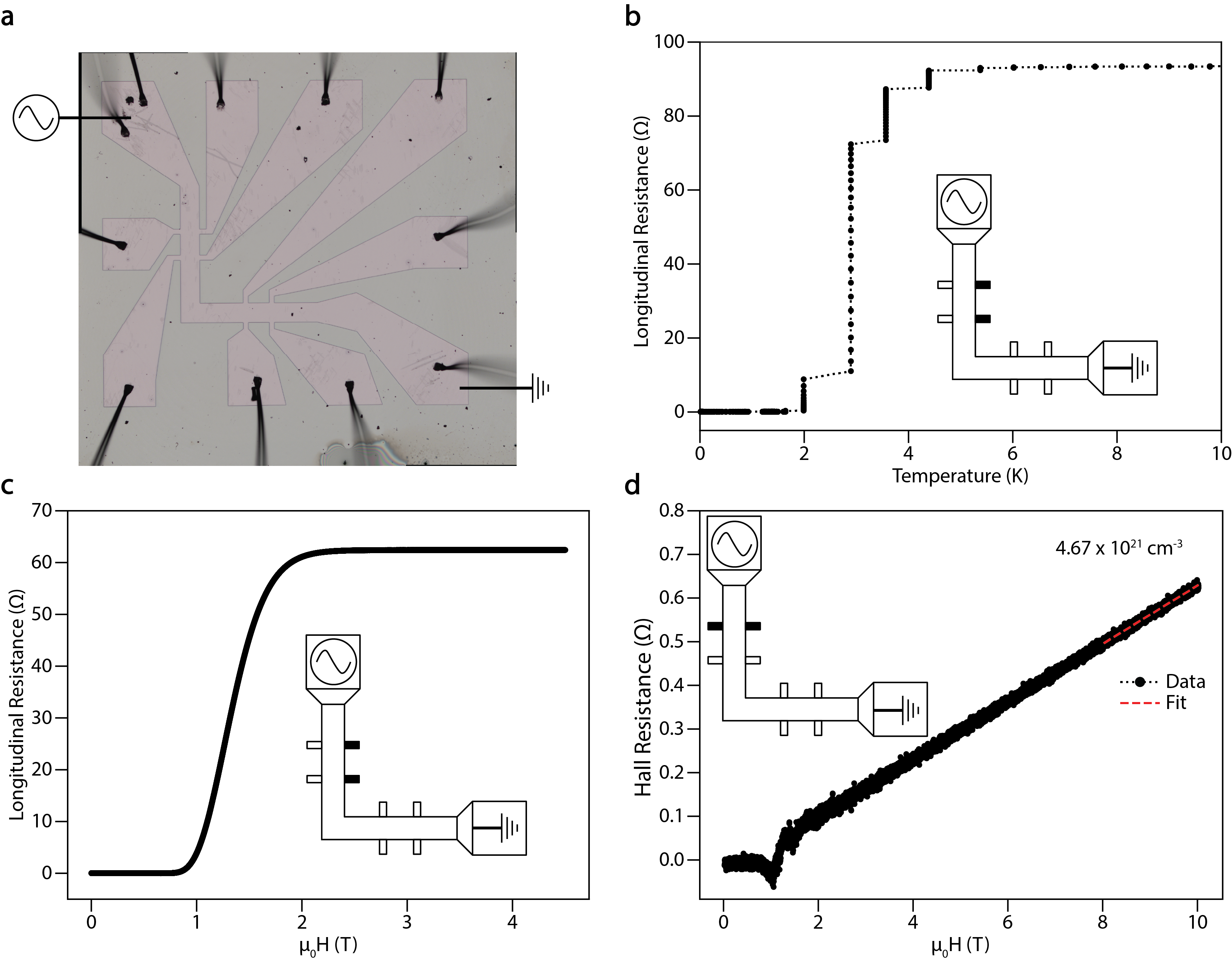}
    \caption{Hall bar device fabrication and transport characterization of a bilayer heterostructure. \textbf{a} Presents an optical image of the fabricated device after wirebonding. We measure under current bias using a lock-in amplifier at low frequency to reduce noise in our measurements. \textbf{b} Shows the longitudinal resistance data upon cooldown of the device. These measurements are taken in a dilution refrigerator to ensure the sample remains superconducting down to mK temperatures with no re-entrant phase. However, we note that dilution fridges are non-ideal for measurements at higher temperatures where we see these vertical line artifacts during our measurement. This is caused by a delay in the thermometry of our fridge in updating the temperature of the mixing chamber plate. The transition temperature exists around 3~K, similar to the Van der Pauw measurement reported in Figure \ref{fig1}c on the same wafer. \textbf{c} Longitudinal resistance as a function of out-of-plane magnetic field also reports a critical magnetic field of the superconducting state to be $\sim$1~T, similar to what is reported for the Van der Pauw device presented in Figure~\ref{S16}a. \textbf{d} Hall trace on this device from which we calculate the carrier concentration to be $4.67 \times 10^{21}$~cm$^{-3}$, which is comparable to the Van der Pauw measurement of $4.15 \times 10^{21}$~cm$^{-3}$. From this data we conclude that the hyperdoped germanium thin film material is highly robust to typical photolithography and chlorine dry etch processing (more details are found in the Methods section). The inset schematic cartoons in \textbf{b}, \textbf{c}, and \textbf{d} represent the specific pins being measured for the presented traces based on the optical image shown in \textbf{a}. The dark boxes represent the pins being measured in these schematics.}
    \label{S5}
\end{figure}

\clearpage
\section{X-ray absorption measurement alignment and fitting}

\begin{figure}[htbp!]
    \centering
    \includegraphics[width=0.6\linewidth]{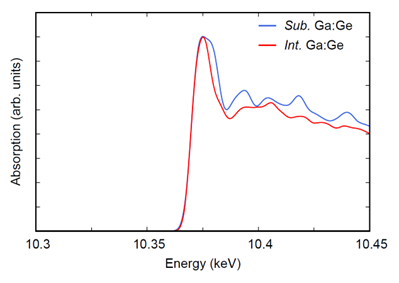}
    \caption{Calculated X-ray absorption spectra of Ga K-edge for Ga:Ge crystals with substitutional and interstitial doping sites.}
    \label{S6}
\end{figure}

\begin{figure}[htbp!]
    \centering
    \includegraphics[width=0.75\linewidth]{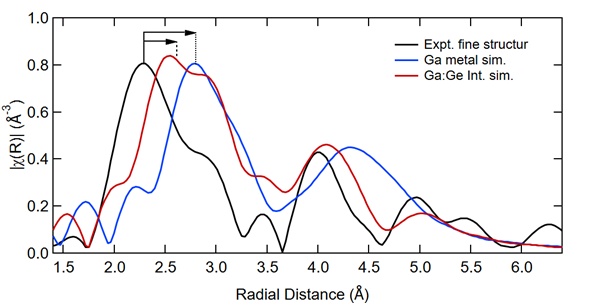}
    \caption{Comparison of the simulated fine structure derived from the CIF files of Ga:Ge interstitial system (DFT calculated system in Fig.~\ref{fig2}a of main text) and amorphous Ga metal~\cite{sharma1962}. Note that these Fourier transforms have been phase corrected, for simplicity. The arrows highlight the relatively large difference in experimental bond length of the nearest-neighbor scattering pair and those derived in the simulations.}
    \label{S7}
\end{figure}

\clearpage
\section{Impact of dopants on electronic structure}

\begin{figure}[htbp!]
    \centering
    \includegraphics[width=\linewidth]{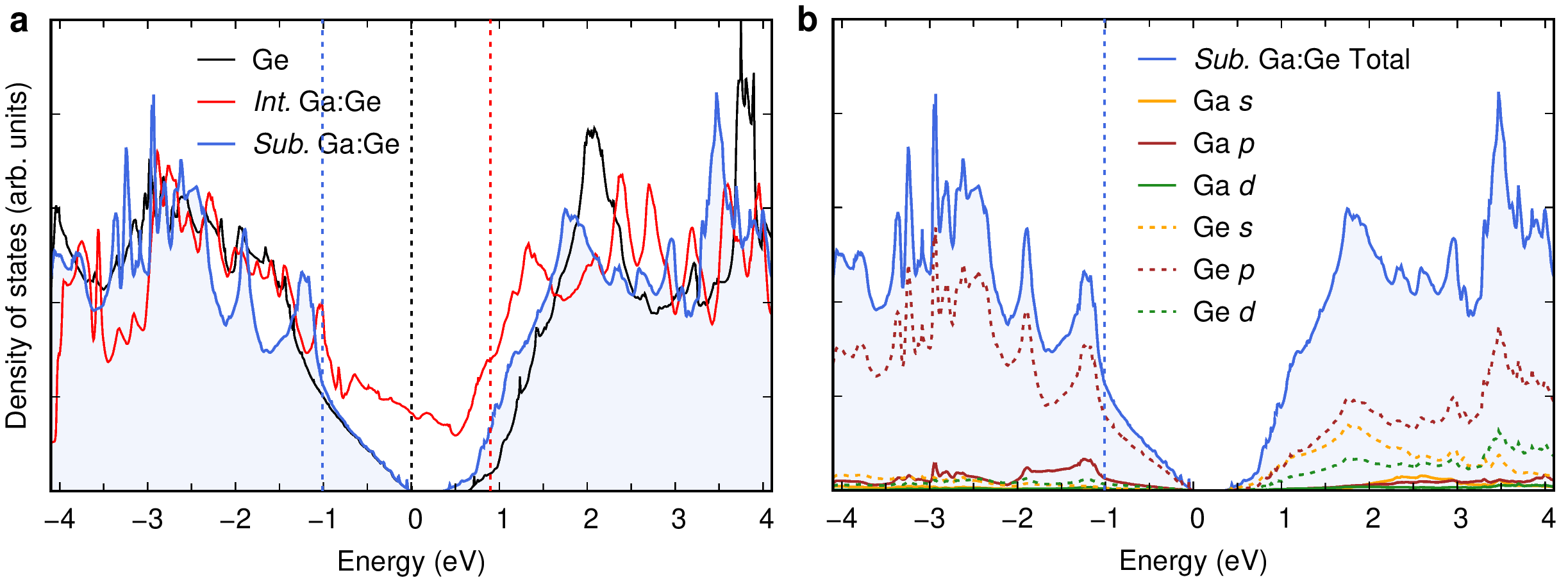}
    \caption{\textbf{a} Total density of states (DOS) calculated for the pure Ge crystal and the hyperdoped Ga:Ge shown in Fig.~\ref{fig2}a of the main text, with Ga at interstitial (int.) and substitutional (sub.) lattice positions. The energies are aligned to the valence band maximum (VBM), set as the zero of the energy. The Fermi level in each case is indicated by a dashed vertical line. We observe further evidence that interstitial doping is unfavorable for for this system as it destroys the ``clean'' electronic bands of the host Ge semiconductor as evidenced by the disappearance of the parent Ge band gap and the emergence of metallic states. In contrast, substitutional Ga doping is expected to form acceptor states below the valence band edge, while the overall band structure near the gap remains largely unperturbed (see Figure~\ref{fig2}d). \textbf{b} Total and projected DOS of the hyperdoped Ga:Ge crystal with substitutional fine-structure.}
    \label{S8}
\end{figure}

\begin{figure}[htbp!]
    \centering
    \includegraphics[width=0.75\linewidth]{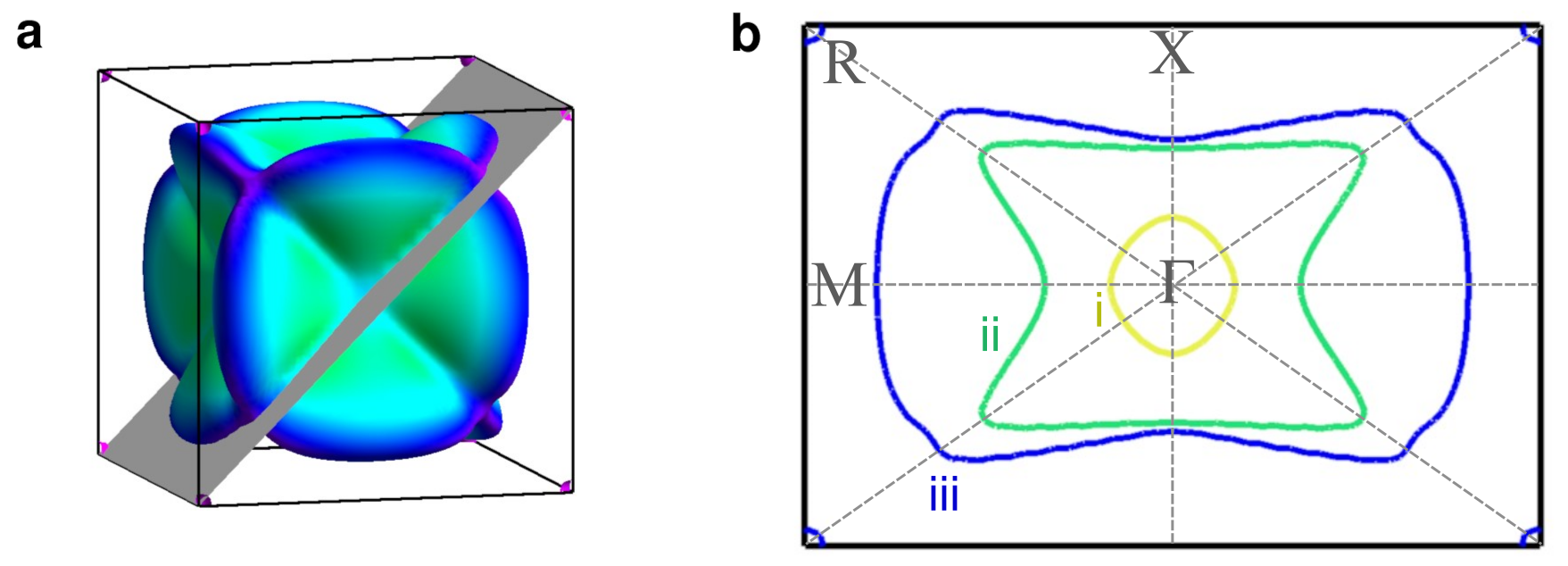}
    \caption{\textbf{a} 3D Fermi surface for substitutionally doped Ga:Ge, showing the (011) plane. \textbf{b} 2D Fermi surface cut on the (011) plane, showing the contribution of bands i, ii, and iii. Hole pockets from band iii are visible in the R-point corners.}
    \label{S9}
\end{figure}

\begin{figure}[htbp!]
    \centering
    \includegraphics[width=0.8\linewidth]{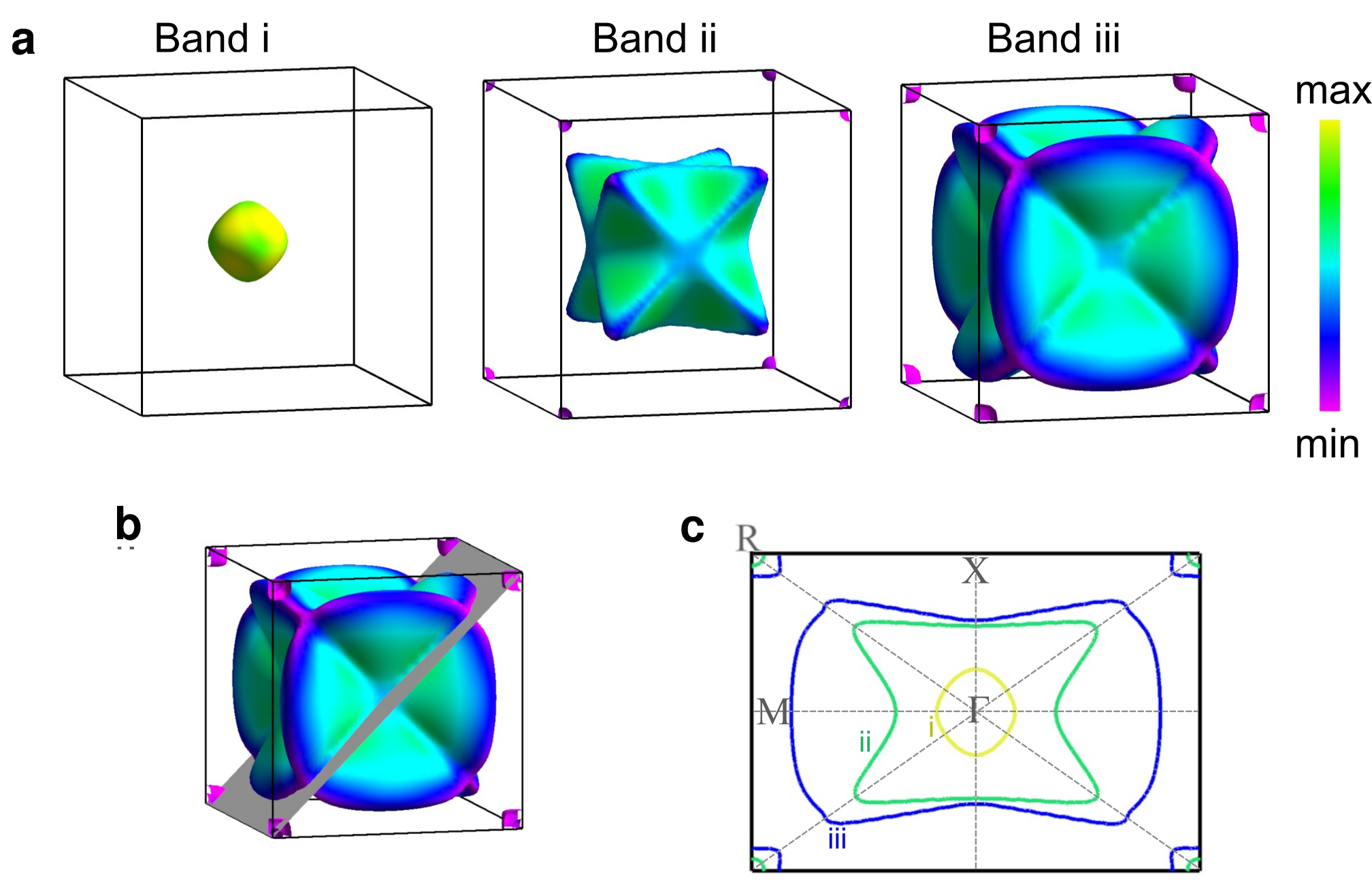}
    \caption{Fermi surface obtained after shifting the Fermi level by 40~meV below the value shown in Fig.~\ref{fig2}d of the main text. \textbf{a} Contribution of bands i, ii, and iii to the 3D Fermi surface, with the color corresponding to the Fermi velocity. \textbf{b} (011) plane crossing the 3D Fermi surface. \textbf{c} 2D cut on this plane, showing the contribution of each band. The predicted narrow- or flat-band conditions are shown to be sensitive to the position of the Fermi level, as the pockets of high-mass carriers from bands ii and iii at the corners of the Brillouin Zone (R point) are shown to enlarge with a relatively minor shift in the Fermi level.}
    \label{S10}
\end{figure}

\clearpage
\section{Domain structure and near-surface Raman spectroscopy}

\begin{figure}[htbp!]
    \centering
    \includegraphics[width=0.8\linewidth]{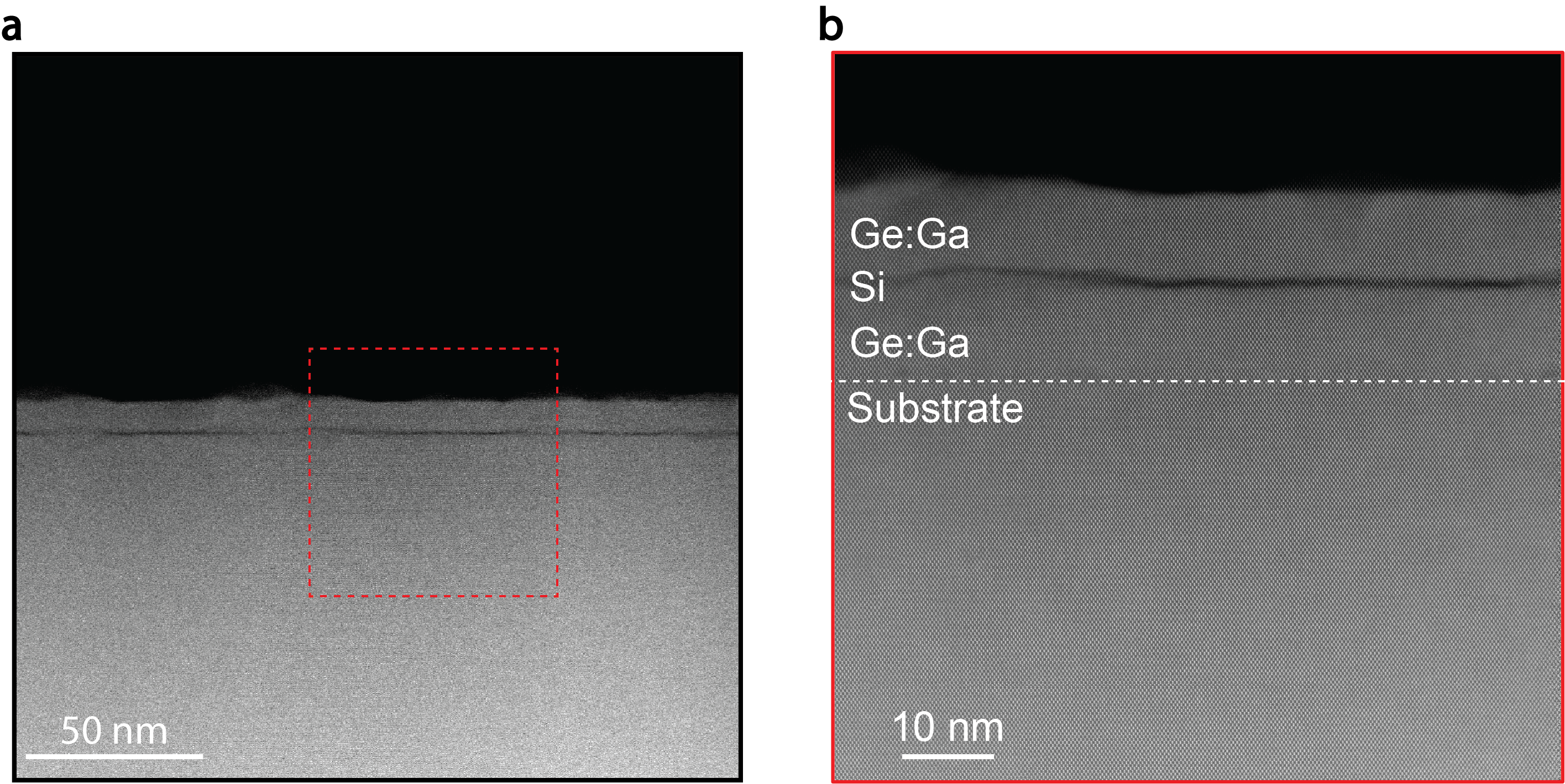}
    \caption{\textbf{a} Low magnification image of the STEM lamella presented in Fig.~\ref{fig3} in the main text. We observe no domain structure as reported in previous reports of MBE-grown hyperdoped Ge films~\cite{strohbeen2023superge}. The red dashed-line box is the region of higher magnification presented in \textbf{b} and Figure~\ref{fig3}d.}
    \label{S11}
\end{figure}

\begin{figure}[htbp!]
    \centering
    \includegraphics[width=0.95\linewidth]{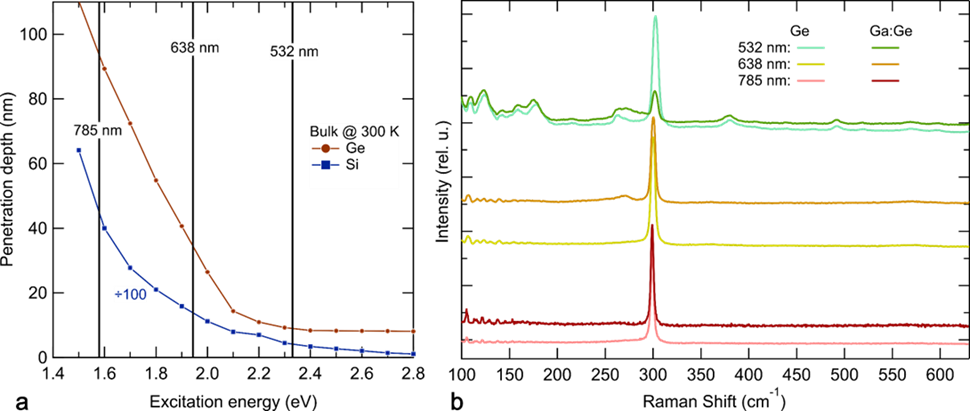}
    \caption{\textbf{a} Raman Scattering penetration depth ($\mathrm{d_\mathrm{RS}}$) as a function of the laser excitation energy, determined by considering the absorption coefficients~\cite{PhysRevB.27.985} for both the incoming ($\alpha_\mathrm{out}$) and the outgoing ($\alpha_\mathrm{in}$) light: $\mathrm{d_\mathrm{RS}} = 1/(\alpha_\mathrm{in} + \alpha_\mathrm{out}) \approx 1/\alpha_{in}$. \textbf{b} Comparison of Raman spectra recorded from Ga:Ge and Ge control MBE films at different excitation energies. At longer excitation wavelengths (785~nm), both the control and target films exhibit a comparable Raman response. Probing the surface-sensitive information with shorter wavelengths, the differences between the vibrational structure of the films become clearer. The appearance of disorder-activated LA modes below 200~cm$^{-1}$ (from both Ge-type and Si-type sublattices) in the scattering volume yields comparable scattering intensities under these conditions for the control and doped sample. We suggest the origins of these bands are a combination of (local) lattice strain, phonon confinement and  (partial) satisfaction of the resonance conditions of the Ge crystal $E_{1}+\Delta_{1}$ gap~\cite{alonso_resonance_1988}. Together, these features break the normal Raman scattering selection rules of a normal (001)-oriented bulk Ge film, i.e., resembling more the Raman spectra recorded using 785~nm laser light.}
    \label{S12}
\end{figure}

\clearpage
\section{Calculation of electron-phonon coupling}

\begin{figure}[htbp!]
    \centering
    \includegraphics[width=0.7\linewidth]{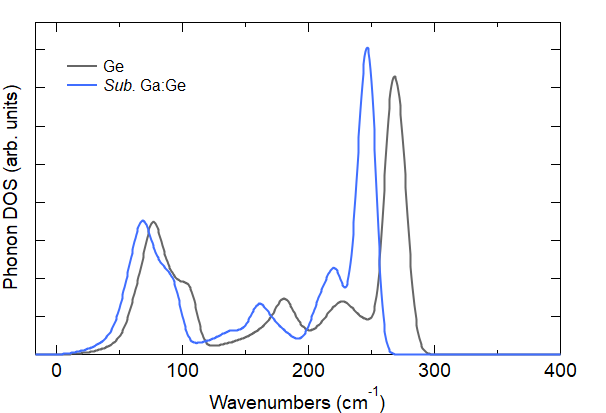}
    \caption{Calculated first-order phonon DOS for pure Ge and hyperdoped Ga:Ge crystal.}
    \label{S13}
\end{figure}

\begin{figure}[htbp!]
    \centering
    \includegraphics[width=0.7\linewidth]{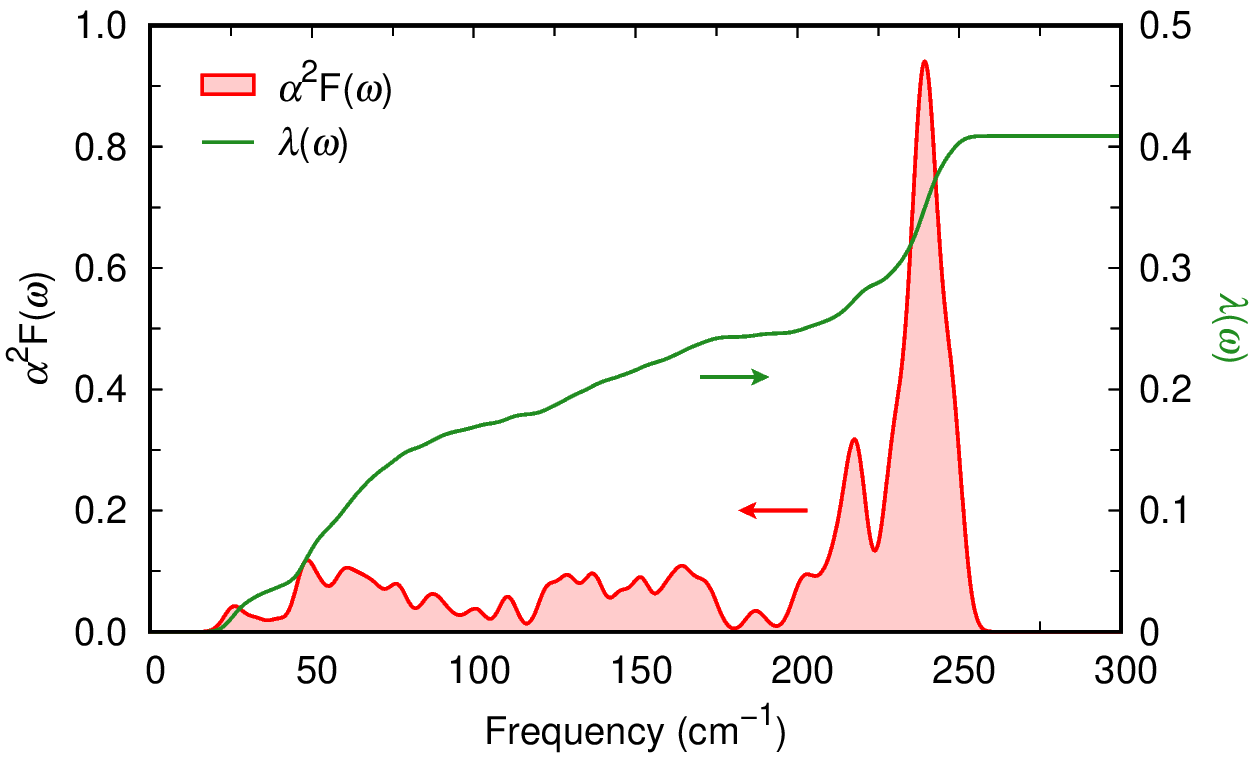}
    \caption{Eliashberg spectral function 
    $\alpha^2F(\omega)$ calculated for substitutionally doped Ga:Ge and cumulative electron-phonon coupling strength $\lambda(\omega)$, showing that the coupling originates mostly from optical phonon modes. 
    The total integrated $\lambda$ is 0.41, yielding a value of 
    $T_\textup{c}=0.77$~K from the McMillan-Allen-Dynes formula [Eq.~\eqref{eq:Tc}].}
    \label{S14}
\end{figure}

To investigate the role of phonons as an underlying mechanism for the observed superconducting state, we perform additional calculations on the electron-phonon coupling strength within our model of hyperdoped Ge. Specifically, we compute the Eliashberg spectral function $\alpha^{2}F$~\cite{Giustino2017}, details of which are reported in the Methods section, to yield the total coupling strength, $\lambda$. From these quantities, the superconducting transition temperature, $T_\textup{c}$, is estimated using the McMillan-Allen-Dynes formula~\cite{AllenDynes1975, McMillan1968}. We calculate $\lambda=0.41$ and $T_\textup{c} = 0.77$~K, providing evidence of a conventional phonon-mediated BCS-type pairing mechanism within this system. The Eliashberg spectral function reported in Figure~\ref{S14} shows that this coupling originates primarily from the optical phonon modes, similar to what has been previously observed in the case of boron-doped diamond~\cite{ekimov2004bcsuper} and silicon~\cite{bustarret2006sib}.\\

\clearpage
\section{Synchrotron penetration depth and calculation of characteristic superconductor length scales}

\begin{figure}[htbp!]
    \centering
    \includegraphics[width=0.75\linewidth]{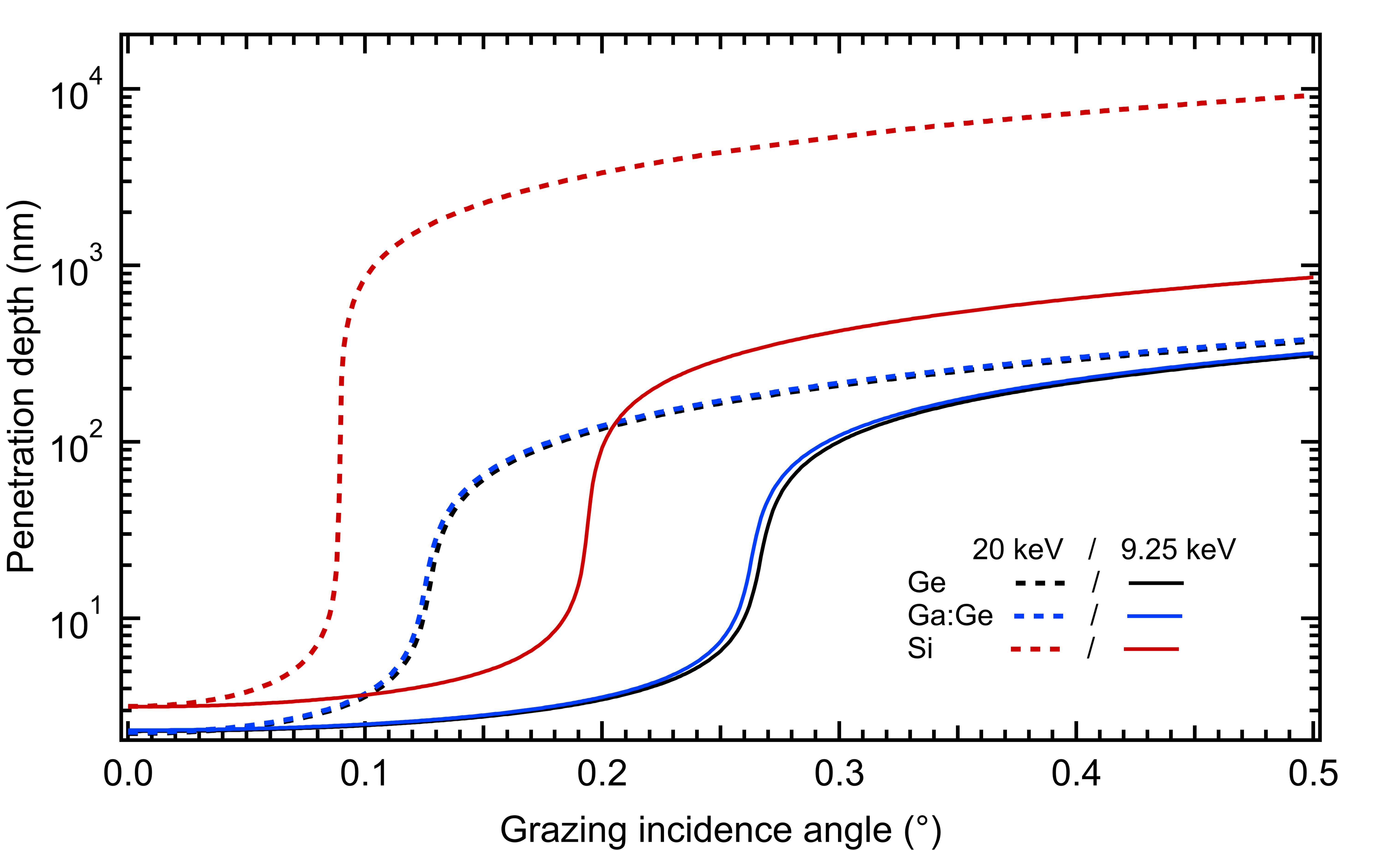}
    \caption{The calculated penetration depth of the different material layers at the synchrotron energies used for grazing incidence experiments. For the range of angles below and at the critical angle, the X-ray electromagnetic field only interacts a short distance below the film surface due to the evanescent damping (5 - 10~nm), and constitutes the defined evanescent regime.}
    \label{S15}
\end{figure}

\begin{figure}[htbp!]
    \centering
    \includegraphics[width=0.8\linewidth]{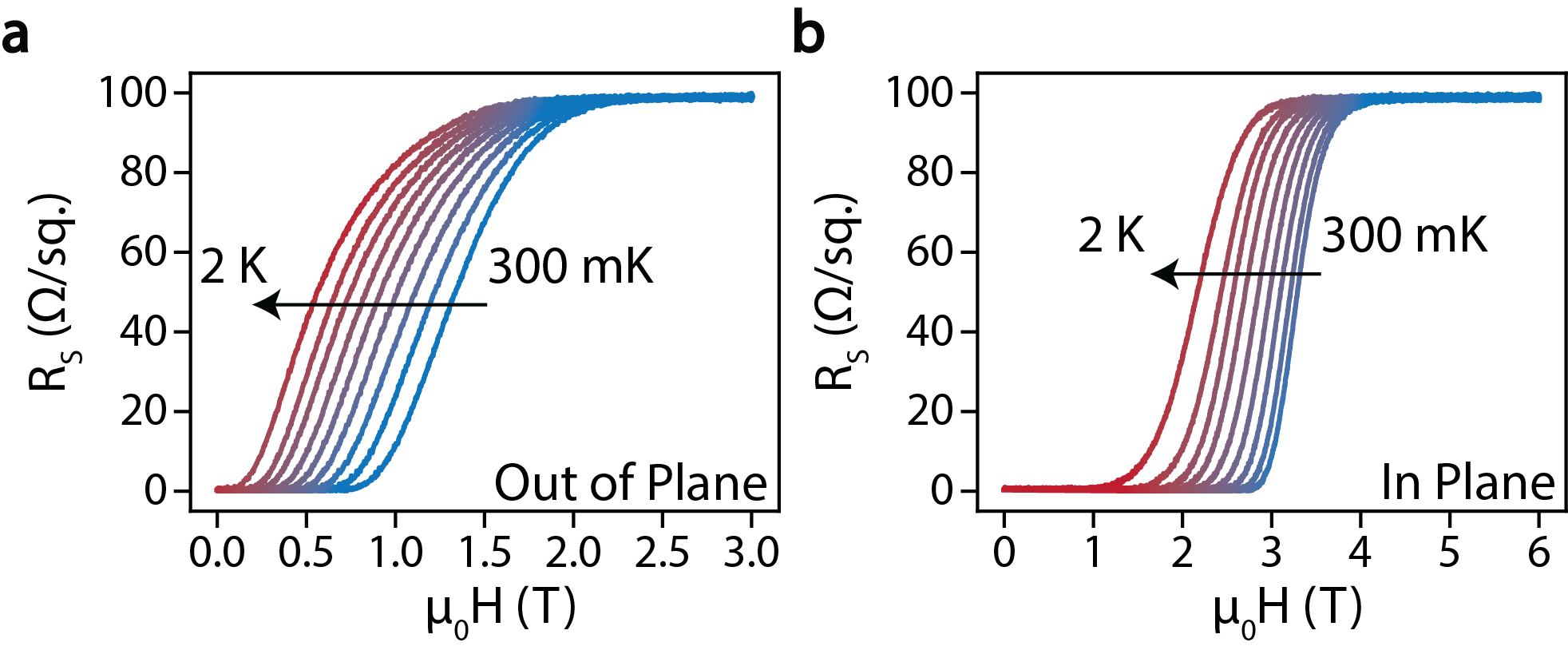}
    \caption{Out-of-plane \textbf{a} and in-plane \textbf{b} critical magnetic field behavior for the Ga:Ge sandwich heterostructure presented in Figure \ref{fig1}c and Figure \ref{fig3}.}
    \label{S16}
\end{figure}

\begin{figure}[htbp!]
    \centering
    \includegraphics[width=0.9\linewidth]{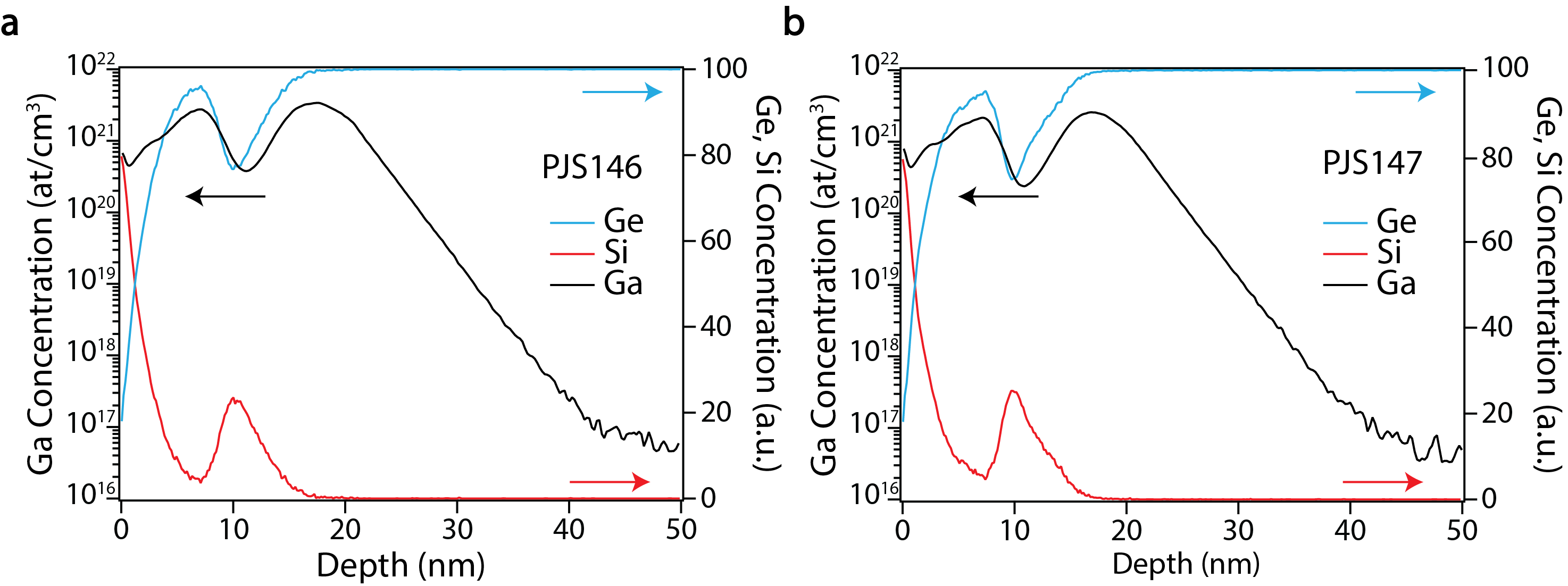}
    \caption{Secondary-Ion Mass Specroscopy data for two samples investigated in this study, \textbf{a} PJS146 ($T_\textup{c} \sim$ 2.5~K) and \textbf{b} PJS147 ($T_\textup{c} \sim$ 3.5~K).}
    \label{S17}
\end{figure}

The magnetic field dependence of longitudinal resistance within the Josephson structure is presented in Figure~\ref{S16}a and b for out-of-plane and in-plane field configurations, respectively. At 300~mK we observe the out-of-plane $H_\textup{c}$ to be 0.79~T, decreasing to 0.14~T at 2~K. In-plane $H_\textup{c}$ is found to be 2.51~T, decreasing to 0.78~T at 2~K. We calculate the coherence length from this data using the following equation:
\begin{equation}
    \xi = \sqrt{\frac{\Phi_\textup{0}}{2 \pi H_\textup{c2}(0)}}
\end{equation}
where $\Phi_\textup{0}$ is the magnetic flux quantum and $H_\textup{c2}(0)$ is the upper critical magnetic field extrapolated to zero Kelvin. London penetration depth is also calculated using:
\begin{equation}
    \lambda_\textup{L} = \sqrt{\frac{m^*}{2 \pi n_\textup{s}q^2}}
\end{equation}
where $m^*$ is the effective mass of the hole carriers which we take to be the weighted average, $0.28m_\textup{e}$, and $n$ is the superfluid density which we obtain from our Hall measurements as $2n_\textup{s} = n_\textup{h}$. We calculate the penetration depth ($\lambda$) and coherence length ($\xi$) on the trilayer sample presented in Figures~\ref{fig3} and \ref{fig4} to be 29~nm and 13.4~nm, respectively, similar to what has been previously observed in boron-doped diamond superconductors~\cite{bustarret2004bccoh}.

\clearpage
\section{GIWAXS alignment, control, and stress analysis}

\begin{figure}[htbp!]
    \centering
    \includegraphics[width=0.75\linewidth]{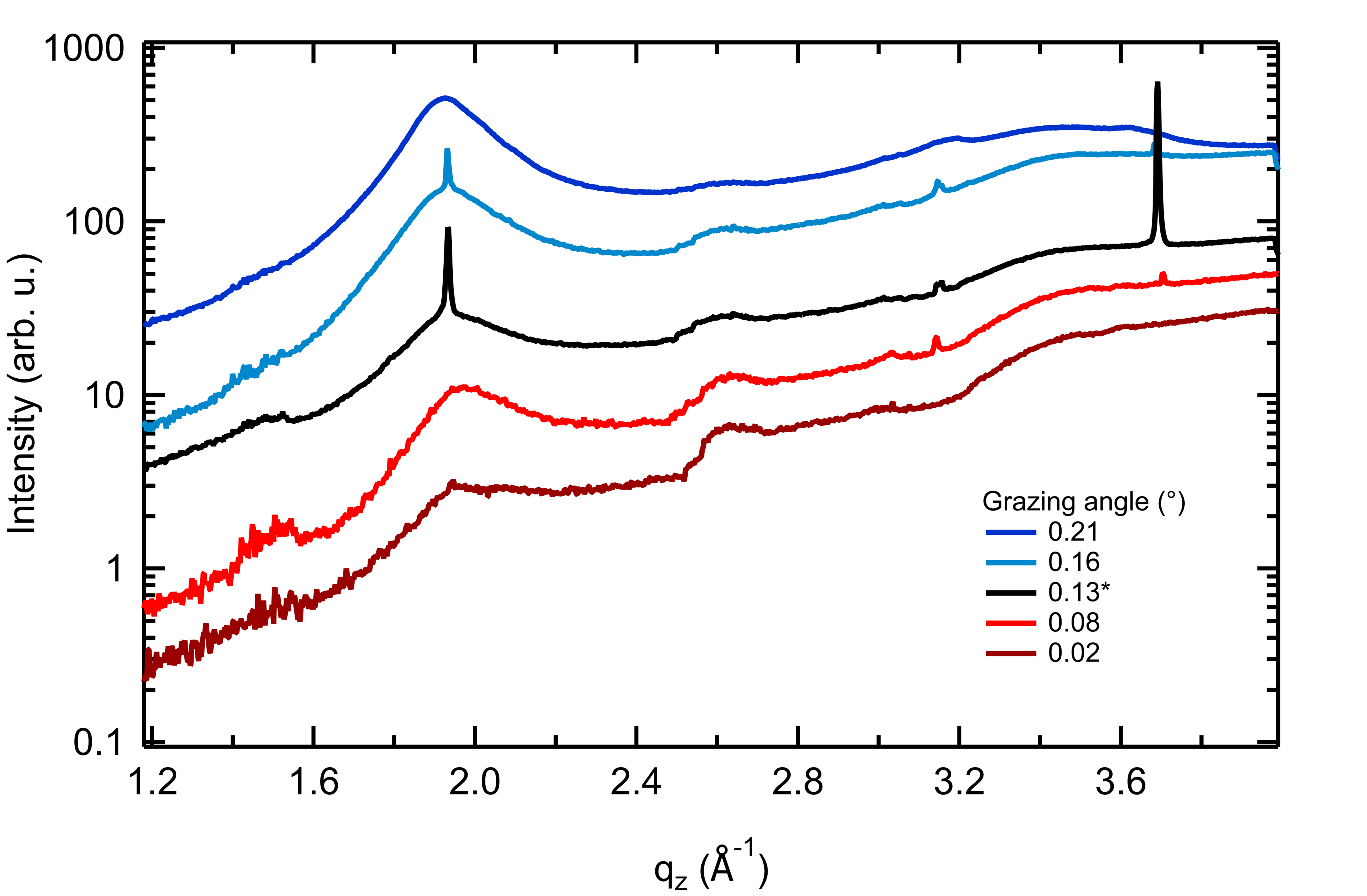}
    \caption{Evolution of the integrated GIWAXS signal recorded from the Ga:Ge film as a function of the grazing incidence angle ($\alpha$i: values inset). Here * identifies the calculated critical angle for the doped Ga:Ge layer at an X-ray beam energy of 20~keV. As the angle of incidence nears the critical angle of the doped layer, the Bragg reflections emerging begin to intensify. A combination of lateral steps (200~$\mu$m) made between each incident angle frame, mechanical jolting, and imperfect beam divergence means the critical angle conditions (waveguide type effect) are relatively sensitive and result in the discontinuity of some profile features between frames.}
    \label{S18}
\end{figure}

\begin{figure}[htbp!]
    \centering
    \includegraphics[width=0.75\linewidth]{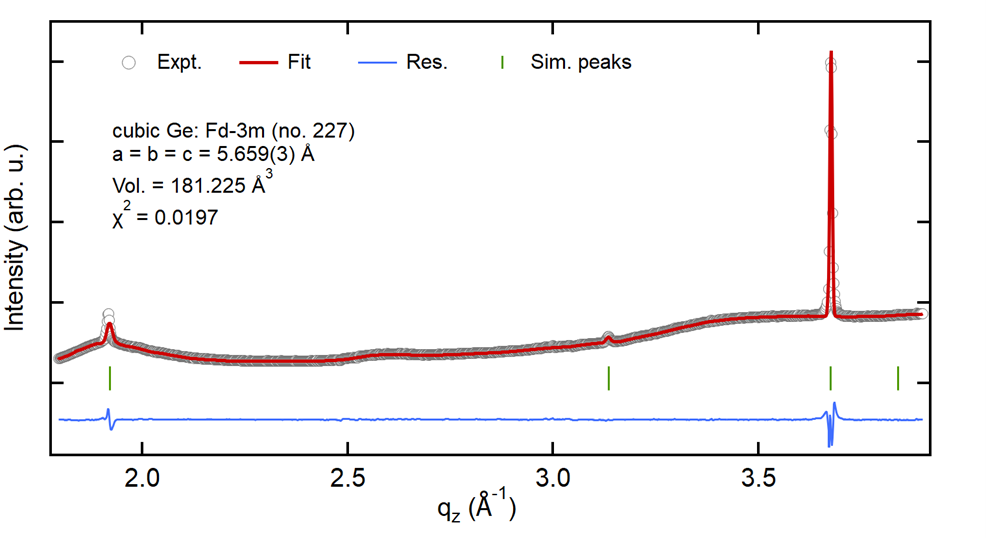}
    \caption{Integrated GIWAXS profile (qxyz) and structural refinement (Le Bail method) of the pure Ge epitaxial layer. Unstrained Ge crystal belongs to the high-symmetry cubic space group $Fd\overline{3}m$ (space group No. 227), characterized by equivalent lattice parameters ($a=b=c$) and a structural refinement of our undoped Si-Ge film aligns well with the reported bulk, cubic value ($a=5.657906\pm0.0000009$~\AA, Vol. = 181.1203~\AA$^{3}$)~\cite{baker1975}.}
    \label{S19}
\end{figure}

\begin{center}
    \textbf{Stress Analysis}
\end{center}
The lattice mismatch is accommodated by an elastic strain in the hyperdoped epilayer, giving a biaxial stress of:
\begin{equation}
    \sigma = 2\mu f \cdot (1 + \nu)/(1 - \nu)    
\end{equation}
where $\mu$ is the shear modulus, $\nu$ Poisson's ratio and $f$ the misfit parameter. Assuming the hyperdoped Ga:Ge epilayer forms a cubic unit cell when unconstrained, elastic isotropy is assumed in the relevant deformation directions and the misfit parameter is given by:
\begin{equation}
    f = \frac{(a_\textup{Ga:Ge} - a_\textup{Ge})}{a_\textup{Ge}}    
\end{equation}
where $a_\textup{Ga:Ge}$ and $a_\textup{Ge}$ are the lattice parameters of the relaxed epilayer and substrate, respectively. We do not have such experimental data for the relaxed hyperdoped unit cell, however, since the epilayers studied here are pseudomorphically grown (Figure~\ref{fig3}e), the values obtained from the structural refinement can be related to the relaxed lattice parameter of the Ga:Ge layer through:
\begin{equation}
    a_\textup{Ga:Ge}=a_\textup{Ge}[1+P(\Delta d/d)_\perp]    
\end{equation}
where $P$ is the elastic parameter of the films~\cite{mcskimin1963}. Regardless of the growth direction, the in-plane and out-of-plane biaxial strain components are given by: 
\begin{equation}
    \epsilon_{\parallel} =(a_\textup{Ga:Ge} - a_\textup{Ge})/a_\textup{Ge}    
\end{equation}
\begin{equation}
    \epsilon_{\perp} = \epsilon_{\parallel}(1 - 1/P)
\end{equation}
respectively. For our thin films grown on Ge (100); $P_{100} = C_{11}/(C_{11}+2C_{12}) = 0.5709$. To implement our calculation, we approximate the elastic constants $C_{ij}$ of the hyperdoped film to be comparable to the pure Ge crystal~\cite{mcskimin1963} (as there are no relevant data for GaGe alloys). Based on our hole carrier concentrations, the Ga concentration for the sample studied in Figure~\ref{fig3}e is $\sim$11.22\% and has a calculated relaxed (cubic) lattice parameter of $a_\textup{Ga:Ge}$ = 5.6614~\r{A}, which is 0.11\% larger than the Ge crystal.\\

\clearpage
\section{DFT and EXAFS data tables}

\begin{table}[htbp!]
    \centering
    \caption{\textbf{Structural parameters of DFT optimized structures and defect formation energies $E_{\textup{Form}}$.} Values are calculated for pure and Ga-doped Ge (both substitutional and interstitial doping).}
    
    \begin{tabular}{ccccc}
    \toprule
        \textbf{Structure} & \textbf{Lattice Parameter (\r{A}}) & \textbf{Volume (\r{A}$^{3}$}) & \textbf{Crystal Structure} & \textbf{$E_{\textup{Form}}$} \textbf{(eV)} \\
        & \textbf{a=b},  \textbf{c} & & & \\
        \midrule
        Pure Ge & 5.674,  5.674 & 182.64 & Cubic & \\
        Sub. Ga:Ge & 5.674,  5.681 & 182.87 & Tetragonal & 0.314 \\
        Int. Ga:Ge & 5.674,  6.220 & 200.128 & Tetragonal & 1.946 \\
        \bottomrule
        \label{dft-table}
    \end{tabular}
\end{table}

\begin{table}[htbp!]
    \centering
    \caption{\textbf{EXAFS fitting parameters.} Values are extracted from fits made to spectra recorded at 295 K using a Ga substitutional model. The lower ranked paths (in italics) are omitted from the actual fit model to avoid overfitting. The errors provided in the parentheses align with the last significant figures of the variable fit values. Further fitting details can be found in the Methods section. The path type can be tracked as: SS - single scattering pair; DS - double scattering; R - rattle.}
    \begin{tabular}{cccccccc}
    \toprule
        \textbf{Path type} & \textbf{Degen.} & \textbf{Rank} & \textbf{R(\r{A})} & \textbf{$\sigma^{3}$ (\r{A}$^{2}$)} & \textbf{$\Delta$E$_{0}$ (eV)} & \textbf{S0$_{2}$} & \textbf{R-factor} \\
        \midrule
        Ga-Ge (SS) & 4 & 100 & 2.451(1) & 0.0051(4) & & &  \\
        Ga-Ge (SS) & 12 & 90.82 & 4.002(5) & 0.0063(2) & & &  \\
        \textit{Ga-Ge-Ge (DS)} & \textit{12} & \textit{5.71} & \textit{4.450} & & & & \\
        Ga-Ge-Ge (DS) & 24 & 21.4 & 4.452(2) & 0.010(6) & -0.195 & 0.68(11) & 0.015 \\
        Ga-Ge (SS) & 12 & 59.45 & 4.693(2) & 0.0087(5) & & & \\
        \textit{Ga-Ge-Ge (R)} & \textit{4} & \textit{4.09} & \textit{4.900} & & & & \\
        \bottomrule
        \label{exafs-table}
\end{tabular}
\end{table}

\section*{Data and materials availability:}
Raw data available on Zenodo: doi.org/10.5281/zenodo.17065133~\cite{zenodo} \\

\end{widetext}
\bibliography{bibliography}

\end{document}